\documentclass[aps,twocolumn,superscriptaddress]{revtex4-2}

\usepackage{blindtext}
\usepackage{amsmath,amssymb,mathrsfs}
\usepackage{natbib}
\usepackage{subfigure}
\usepackage{tabularx}
\usepackage{epsfig}
\usepackage{longtable}
\usepackage{amsfonts}
\usepackage{rotating}
\usepackage{subfigure}
\usepackage{amsmath}
\usepackage{comment}
\usepackage{bbold}
\usepackage{color}
\usepackage{braket}
\usepackage{svg}
\usepackage{verbatim}
\usepackage{float} 
\usepackage{booktabs} 
\usepackage{graphicx} 
\usepackage{amsmath}
\usepackage{array}
\usepackage{mathtools}
\usepackage{physics}
\usepackage{bm}

\usepackage[colorlinks=true]{hyperref}

\hypersetup{linkcolor=blue,citecolor=blue,urlcolor=blue}

\def\be{\begin{equation}}
\def\ee{\end{equation}}
\def\bea{\begin{eqnarray}}
\def\eea{\end{eqnarray}}

\begin{document}

\title{Topology-ferrimagnetism intertwining via weak interactions in Lieb lattices}

\author{Lei Chen}
\thanks{These authors contributed equally.}
 \affiliation{Department of Physics, National University of Defense Technology, Changsha 410073, China}
\author{Bei-Bei Wang}
\thanks{These authors contributed equally.}
\affiliation{School of Physics and Institute for Quantum Science and Engineering, Huazhong University of Science and Technology, Wuhan 430074, China}
 \author{Jianmin Yuan}
  \affiliation{Institute of Atomic and Molecular Physics, Jilin University, Changchun 130012, China}
 \affiliation{Department of Physics, National University of Defense Technology, Changsha 410073, China}
\author{Long Zhang}
 \email{lzhangphys@hust.edu.cn}
 \affiliation{School of Physics and Institute for Quantum Science and Engineering, Huazhong University of Science and Technology, Wuhan 430074, China}
 \affiliation{Hefei National Laboratory, Hefei 230088, China}

\author{Jinsen Han}
\email{hanjinsen12@nudt.edu.cn}
\affiliation{Department of Physics, National University of Defense Technology, Changsha 410073, China}
\affiliation{Hunan Key Laboratory of Extreme Matter and Applications, National University of Defense Technology, Changsha 410073, China}
\affiliation{Hunan Research Center of the Basic Discipline for Physical States, National University of Defense Technology, Changsha 410073, China}
\author{Yongqiang Li}
 \email{li\_yq@nudt.edu.cn}
 \affiliation{Department of Physics, National University of Defense Technology, Changsha 410073, China}
 \affiliation{Hunan Key Laboratory of Extreme Matter and Applications, National University of Defense Technology, Changsha 410073, China} 
 \affiliation{Hunan Research Center of the Basic Discipline for Physical States, National University of Defense Technology, Changsha 410073, China}

\begin{abstract}
A common wisdom about quantum many-body systems is that emergent phases typically fall into either the Landau-Ginzburg paradigm or topological classifications. Experimentally realizing the intertwined emergence of spontaneous symmetry breaking and topological order remains challenging. Here, we present an experimentally accessible platform for studying magnetic topological states in a spin-orbit-coupled Lieb lattice. Remarkably, we observe the coexistence of topological characteristics, quantified by the Chern number and Bott index, with spontaneous symmetry-breaking orders, such as ferrimagnetism, in the many-body ground states. Computational analyses combining dynamical mean-field theory and Hartree-Fock approximations reveal a pronounced parameter regime where magnetic topological insulators emerge even under weak interactions. This unconventional phenomenon originates from the Lieb lattice's unique band structure, which facilitates the synergy between interaction-driven symmetry breaking and spin-orbit coupling induced band inversion. Crucially, spin polarization and spin winding co-emerge as inherently coupled phenomena due to their shared origin in the same interacting, spinful atoms. We further propose a specific experimental implementation scheme for ultracold atoms, utilizing currently available Raman lattice techniques. Our findings pave the way for exploring the interplay between symmetry-broken states and topological order in strongly correlated systems.

\end{abstract}

\date{\today}

\maketitle

{\it Introduction}.
One of the biggest challenges in quantum many-body physics is the classification of emergent phenomena associated with various quantum phases, which forms a cornerstone of modern condensed matter physics~\cite{sachdev1999quantum}. Generally, 
there are two paradigms to classify quantum many-body phases. The first is Landau's symmetry-breaking theory which characterizes phases through local order parameters~\cite{2000SYMMETRY}, including quantum magnetism, superconductivity, and superfluidity~\cite{annett2004superconductivity}. The other is topological classification, which describes quantum phases with nonlocal topological invariants beyond symmetry-breaking descriptions~\cite{qi2011topological,hasan2010colloquium}. In this regard, it is widely accepted that spontaneous symmetry-breaking orders are incompatible with topological phases within the same degrees of freedom~\cite{liu2016quantum,tokura2019magnetic,bernevig2022progress}, while topological phases are typically associated with systems characterized by the absence of order parameters.
To overcome this limitation, a prevalent approach in condensed matter physics requires introducing interactions to investigate magnetic topological states, despite the conventional decoupling of magnetic and electronic degrees of freedom~\cite{li2019intrinsic,otrokov2019prediction,serlin2020intrinsic}.

Ultracold fermionic atoms in optical lattices are ideal candidates for the realization of both topological and Landau's symmetry-broken phases, due to the tunability of various parameters~\cite{Lewenstein_2007,bloch2008many,Esslinger_2010,Dutta_2015,RevModPhys.83.1523,RevModPhys.91.015005}. 
Unfortunately, the intertwined emergence of topology and magnetism has yet to be achieved experimentally 
in ultracold-atom quantum simulators. A main challenge is that magnetic long-range order generally requires strong interactions~\cite{mazurenko2017cold,shao2024antiferromagnetic}, while experimentally achieved nontrivial band topology survives only in weakly interacting regimes~\cite{lin2011spin,PhysRevLett.109.095301,PhysRevLett.109.095302,Jotzu_2014,wu2016realization,Kolkowitz_2016,huang2016experimental,PhysRevLett.117.220401,sun2018highly,Song_2018,Tarnowski_2019,Wintersperger_2020,PhysRevResearch.5.L012006}.
These competing energy scales traditionally preclude simultaneous realization in quantum many-body systems.
However, a recent experimental breakthrough in Lieb lattice systems~\cite{lebrat2024ferrimagnetism} has conclusively validated Lieb's conjectured ferrimagnetic ordering under infinitesimal onsite interactions~\cite{lieb1989two,nguyen2016dynamical}. This advance raises a critical question: Can Lieb lattices simultaneously host ferrimagnetism and nontrivial topology through the introduction of spin-orbit coupling (SOC)?

In this Letter, we address the above pivotal question
by investigating magnetic topological phases in a spin-orbit-coupled Lieb lattice. Through combined theoretical and numerical analyses, we demonstrate that weak interactions suffice to induce both nontrivial topological invariants (Chern number or Bott index) and long-range ferrimagnetic order with staggered spin textures. Unlike prior approaches requiring strong correlations~\cite{PhysRevLett.116.225305,PhysRevLett.120.157205,PhysRevB.103.155108,PhysRevB.106.205107} or those yielding trivial topology at half-filling~\cite{PhysRevB.104.235115}, our scheme enables continuous parameter tuning across weak-to-moderate interaction regimes, where magnetic orders and nontrivial topology can coexist. We further propose a realistic cold-atom implementation of the model using optical Raman lattices~\cite{wu2016realization,song2019observation,PhysRevResearch.5.L012006}, establishing a versatile platform to probe the interplay of spontaneous symmetry breaking and topology.

\begin{figure}[!tb] 
\centering
\includegraphics[width=0.48\textwidth]{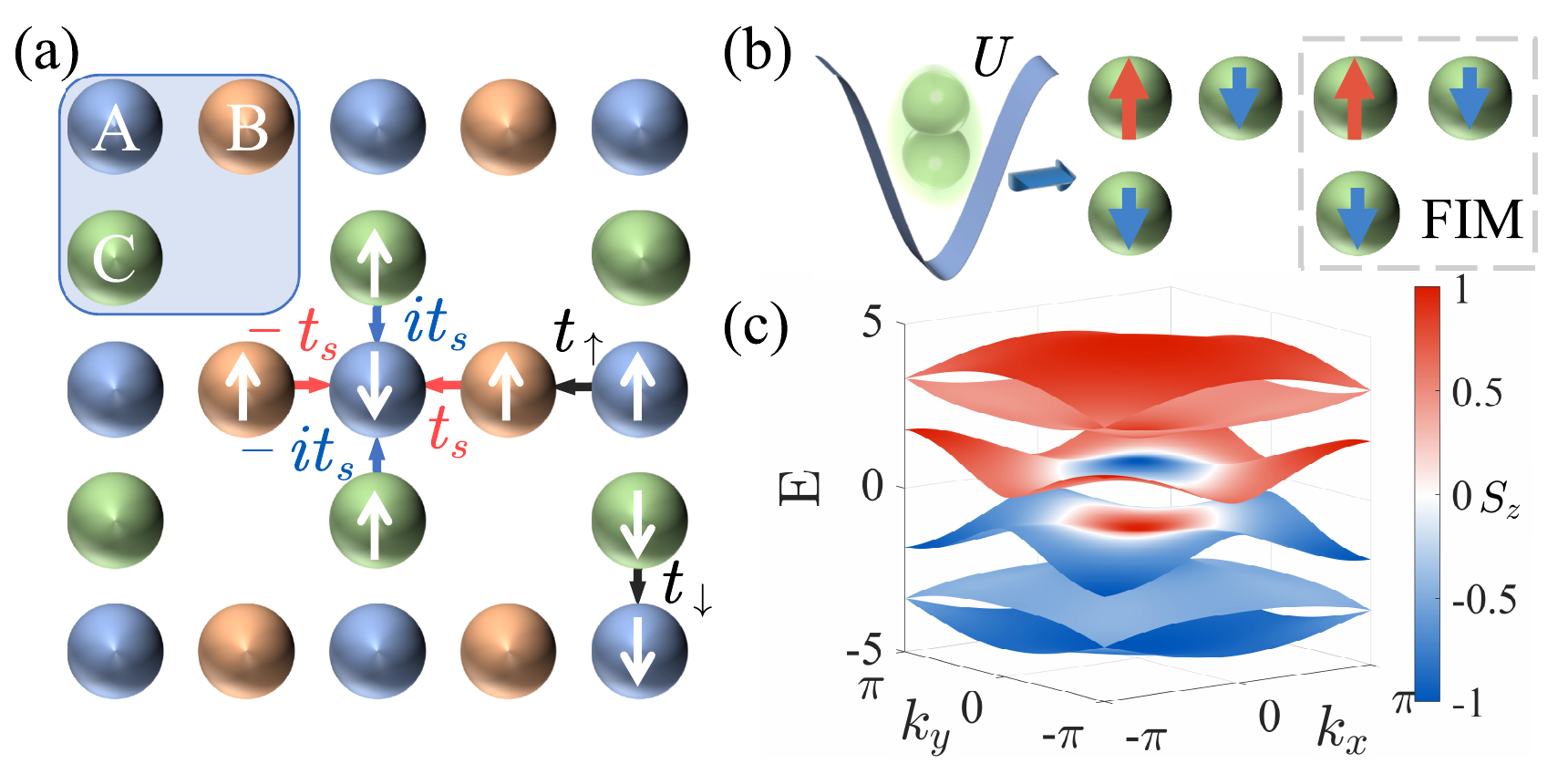} 
  \caption{
  (a) Sketch of the geometrical structure of a 2D spin-orbit-coupled Lieb lattice. The shaded region indicates a unit cell containing three irreducible sites $A$, $B$ and $C$. The hopping amplitudes are denoted as: $t_\uparrow$ and $t_\downarrow$ for spin-conserved hopping, and $t_s$ for spin-flip hopping. The phases associated with these hoppings along different directions ($\pm1$ and $\pm i$) are also depicted.
   It is expected that (b) long-range spin order develops due to the interaction $U$, minimizing the total energy in the Lieb lattice; and (c) nontrivial band topology emerges, induced by SOC parameterized by $t_s$.}
  \label{fig:lieb_model}
\end{figure}

{\it Model and method}. We investigate two-component spin-orbit-coupled fermionic gases in a two-dimensional (2D) Lieb lattice, as shown in Fig.~\ref{fig:lieb_model}(a). 
In a sufficiently deep lattice, the system is well-described by an extended Fermi-Hubbard model, with the Hamiltonian

\begin{equation}
  \begin{split}
      H=
&\sum_{\langle\bf{r},\bf{r}'\rangle,\nu\neq\nu'}t_{s}^{\bf{r}-\bf{r}'}c_{\bf{r},\nu}^{\dagger}c_{\bf{r}',\nu'}-\sum_{\langle\bf{r},\bf{r}'\rangle,\nu}t_{\nu}c_{\bf{r},\nu}^{\dagger}c_{\bf{r}',\nu}\\
&+\sum_{\bf{r},\nu\neq\nu'}	\frac{U}{2} n_{\bf{r},\nu}n_{\bf{r},\nu'}+\sum_{\bf{r}}m_{z}(n_{\bf{r},\uparrow}-n_{\bf{r},\downarrow})\\
&-\sum_{\bf{r},\nu} \mu_{\nu}n_{\bf{r},\nu},
  \end{split}
	\label{eq:Ham}
\end{equation}
where $c_{\bf{r},\nu}^{\dagger}$ ($c_{\bf{r},\nu}$) is the fermionic creation (annihilation) operator for spin $\nu=\uparrow, \downarrow$ at site $\bf{r}$, $t_\nu$ ($t_s$) presents the nearest-neighbor spin-conserved (spin-flip) hopping, $\mu_\nu$ is the chemical potential, $U$ denotes the onsite interaction, $m_z$ is the magnetic field, and $n_{\bf{r},\nu}$ is the particle density.
Here, $\bf{r}$ labels the lattice sites of sublattices $A, B$ and $C$ within the unit cell, and the superscript $\bf{r}-\bf{r}'$ indicates the direction-dependent hopping amplitudes between distinct sublattices, as depicted in Fig.~\ref{fig:lieb_model}(a). 
A typical feature of the Lieb lattice is its bipartite structure with inequivalent sublattices, which hosts a non-dispersive flat band~\cite{flannigan2021hubbard,cui2020realization}. At half-filling, Lieb's theorem rigorously establishes that an infinitesimal interaction strength induces a ground state with nonzero spin per unit cell, manifesting as ferrimagnetism~\cite{lieb1989two,noda2014flat}; see Fig.~\ref{fig:lieb_model}(b). Conversely, introducing SOC alone imbues the system with nontrivial topology, generating band inversion as shown in Fig.~\ref{fig:lieb_model}(c)~\cite{tsai2015interaction,jiang2019topological,banerjee2021higher}. Within this framework, we investigate the interplay of magnetic and topological phases by simultaneously incorporating SOC and onsite interactions.
Throughout the entire article, we set $t_{\uparrow}=-t_{\downarrow} =t_0$.

For investigating the many-body system and characterizing interaction effects, we implement real-space dynamical mean-field theory (DMFT)~\cite{RevModPhys.68.13,PhysRevLett.100.056403,Snoek_2008,PhysRevA.110.043319}, a nonperturbative framework valid across the entire coupling strength  range. Complementary to this approach, the Hartree-Fock (HF) method~\cite{das2024realizing,soni2021multitude} is employed, providing reliable approximations specifically in weak interaction regimes.
Technical details for both methods are documented in the Supplementary Materials (SM)~\cite{SM}.

\begin{figure}[!tb]
  \centering
\includegraphics[width=0.48\textwidth]{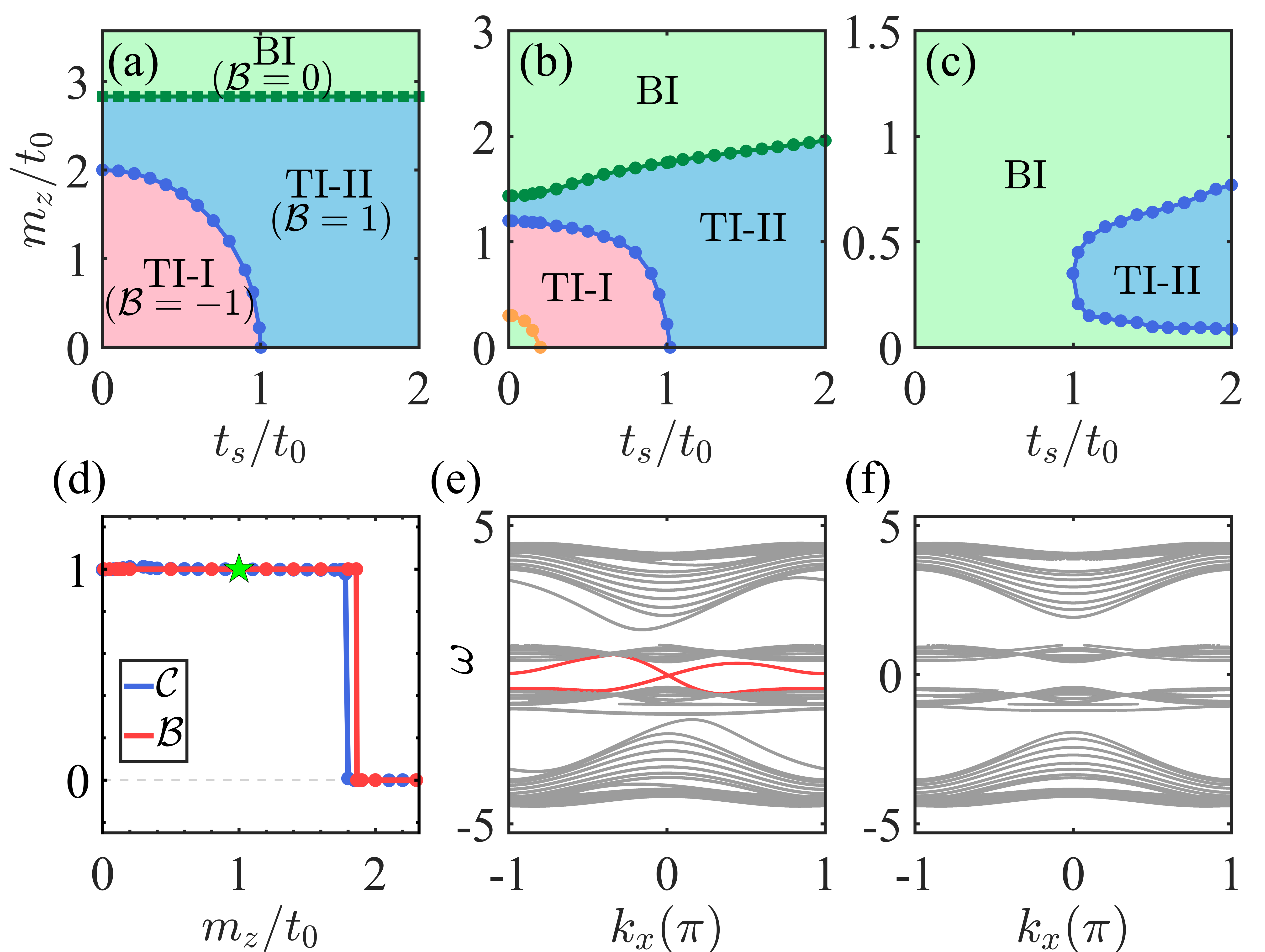} 
  \caption{Topological phase diagrams for free and interacting fermions at half-filling with (a) $U=0$, (b) $U=3t_{0}$, and (c) $U=8t_{0}$, calculated via DMFT. Three distinct quantum phases are identified: a topologically trivial insulator (BI) with Bott index $\mathcal{B} = 0$, and two nontrivial phases [TI-I ($\mathcal{B}=-1$) and TI-II ($\mathcal{B}=1$)]. (d) Chern number $\mathcal{C}$ from HF and Bott index $\mathcal{B}$ from DMFT for the ground state at $U = 3t_{0}$ and $t_s = 1.5 t_{0}$. Spectral function under (e) open, and (f) periodic boundary conditions at $m_z=1.0 t_{0}$ [green star in (d)], obtained by DMFT.}
  \label{fig:topo_phases}
\end{figure}

{\it Topological Phases transition}. We summarize the topological ground-state phase diagrams in Fig.~\ref{fig:topo_phases}(a-c), plotted as functions of $t_s/t_0$ and $ m_z/t_0$ for different interaction strengths. For $U=0$ [Fig.~\ref{fig:topo_phases}(a)], as expected, the system exhibits topologically insulating states when $t_s$ and an appropriate $m_z$ are present, characterized by a non-zero Chern number $\mathcal{C}$~\cite{PhysRevB.87.174402,PhysRevApplied.9.024029,PhysRevB.87.174427} or Bott index $\mathcal{B}$~\cite{han2023interaction,PhysRevLett.125.217202,huang2018theory} and the presence of topological edge states~\cite{PhysRevLett.84.522,PhysRevB.94.214510,vasic2015chiral} (detailed definitions provided in SM~\cite{SM}). Notably, we observe two distinct topological phases: TI-I ($\mathcal{B}= -1$, pink) and TI-II ($\mathcal{B}= 1$, blue). At larger $m_z$, the system transitions to a band-insulating (BI) phase with $\mathcal{B}=0$ (green). 

In the presence of interactions $U$, the topological regions persist but shrink, as demonstrated in Figs.~\ref{fig:topo_phases}(b) with $U=3t_0$ and (c) with $U=8t_0$. The interaction effects are taken into account via the local effective self-energy, which is decomposed as $\Sigma_{\mathrm{eff}}=\frac{\Sigma_{\uparrow\uparrow}-\Sigma_{\downarrow\downarrow}}{2} \sigma_z+\frac{\Sigma_{\uparrow\uparrow}+\Sigma_{\downarrow\downarrow}}{2} \sigma_0 $, where $\Sigma_{\nu\nu}$ is the self-energy for spin component $\nu$, $\sigma_{z}$ is the $z$-component of Pauli matrices, and $\sigma_0$ is the identity matrix~\cite{SM}. The resulting terms act as an effective Zeeman field and a chemical potential, with the former one directly modifying topological properties of the interacting system.
Remarkably, at $U=3t_0$ [Fig.~\ref{fig:topo_phases}(b)], topological phases still occupy a substantial parameter space, demonstrating the stability against interactions. 
In contrast to 2D square lattices where weak interactions destroy topology~\cite{rachel2018interacting,PhysRevLett.109.205303}, the Lieb lattice retains topological phases even in the strongly interacting regime. As shown in Fig.~\ref{fig:topo_phases}(c) with $U=8t_0$, the TI-II phase persists across a pronounced region at strong SOC ($t_s/t_0 \approx 1$).

The topological phase transition at $U=3t_0$ is shown in Fig.~\ref{fig:topo_phases}(d), where nonzero topological invariants (Bott index ${\cal B}$ from DMFT and Chern number 
${\cal C}$ from HF) characterize the half-filled ground state. To further probe the many-body topology, we compute spectral functions via DMFT under cylindrical boundary conditions.
Figs.~\ref{fig:topo_phases}(e) and (f) reveal topological edge states through gap-spanning spectral weight exclusive to open boundaries (e), contrasting with bulk-dominated spectra under periodic boundaries (f).

\begin{figure}[!tb]
  \centering
\includegraphics[width=0.47\textwidth]{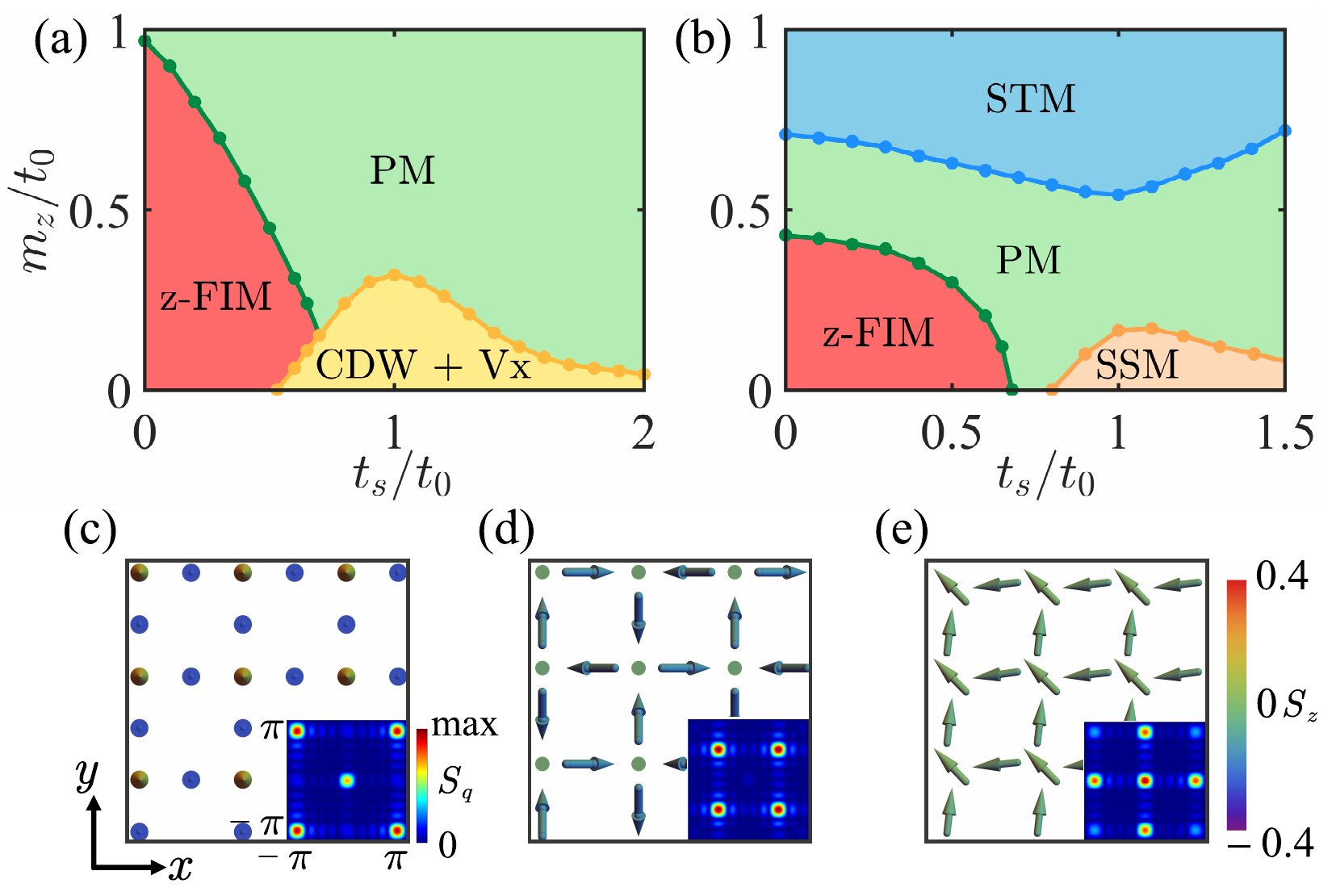} 
  \caption{Magnetic phases at half-filling. Magnetic phase diagrams resolved by DMFT for (a) $U=3t_{0}$ and (b)  $U=8t_{0}$, revealing five distinct phases: ferrimagnetic ($z$-FIM), vortex (Vx), spin-spiral (SSM), partially (PM), and saturated magnetic (STM) phases. (c)-(e) Real-space spin distributions $\langle\mathbf{S}_{\bf r}\rangle$ and momentum-space structure factors $\mathbf{S}_{\bf q}$ (inset) for the $z$-FIM (c), Vx (d), and SSM (e) phases, respectively.}
  \label{fig:mag_order}
\end{figure}

{\it Interaction induced Magnetism}. 
One typical feature of the Lieb lattice is that any infinitesimal
onsite interaction gives rise to a ground state with finite total spin at half filling, $i.e.$ ferrimagnetism. In the absence of SOC, we compute the many-body phase diagram of the interacting fermions in the 2D Lieb lattice, and observe the emergence of ferrimagnetism under symmetry-breaking magnetic field $m_z$~\cite{SM}, consistent with recent experimental observations~\cite{lebrat2024ferrimagnetism}. Further complexity arises with the inclusion of SOC ($t_s$), which drives the emergence of exotic spin-ordered phases.

We present phase diagrams in Figs.~\ref{fig:mag_order}(a) for $U=3t_0$ and (b) for $U=8t_0$, where distinct phases are classified by the spin order parameter $\langle\mathbf{S}_{\bf r}\rangle\equiv \langle c^\dagger_{{\bf r},\nu}{\bm \sigma}_{\nu\nu^\prime}c_{{\bf r},\nu^\prime}\rangle$ (${\bm \sigma}_{\nu\nu^\prime}$ denoting Pauli matrices), spin structure factor $\mathbf{S}_{\bf q} = 1/N\left|\sum_{\bf r} \langle\mathbf{S}_{\bf r}\rangle e^{i\mathbf{q} \cdot \mathbf{r}}\right|$ ($N$ being the number of unit cells), and charge modulation amplitude~\cite{SM}. 
Under weak magnetic field ($m_z < t_0$) and SOC ($t_s < t_0$), the system stabilizes a ferrimagnetic ($z$-FIM) phase characterized by sharp maxima in $\mathbf{S}_{\bf q}$ at momenta $\mathbf{q} = (\pm \pi, \pm \pi)$ [Fig.~\ref{fig:mag_order}(c)], reflecting antiferromagnetic order of $\langle S^z_{\bf r}\rangle$ between $A$ and $B$$C$ 
sublattices.
Crucially, antiferromagnetic correlations inherent to the repulsive Hubbard model competes with $m_z$ (inducing spin polarization) and $t_s$ (promoting spin winding). As shown in Fig.~\ref{fig:mag_order}(a), 
increasing $m_z$ drives a crossover to a 
partially magnetic (PM) phase, signaled by $\langle S^z_{\bf r}\rangle$ sign reversal at the $A$ sublattice and by $\mathbf{S}_{\bf q}$ condensating at $\mathbf{q} = (0,0)$, while enhanced $t_s$ induces a transition to a vortex (Vx) phase with in-plane spin textures and ${\bf S}_{\bf q}$ peaking at $\mathbf{q} = (\pm \pi/2, \pm \pi/2)$ [Fig~\ref{fig:mag_order}(d)].

For strong interaction $U=8t_0$ [Fig.~\ref{fig:mag_order}(b)], 
analogous parameter competition yields distinct phases: larger $m_z$ generates a saturated ferromagnetic (STM) phase 
[${\bf S}_{\bf q}$ peak at $\mathbf{q} = (0,0)$], while larger $t_s$ produces a spin-spiral (SSM) phase [Fig.~\ref{fig:mag_order}(e)].

While magnetic phases in the weakly interacting regime are well understood with the HF mean-field approximation~\cite{SM}, the underlying physics of the strongly interacting regime, such as $U = 8t_{0}$ in Fig.~\ref{fig:mag_order}(b), can be given by an effective spin-exchange model where the hoppings act as perturbations to the dominant interactions~\cite{SM}:
\begin{equation}
\begin{split}
H_{\mathrm{eff}}=
&\sum_{\left\langle\mathrm{r,r'}\right\rangle}[J^x_{\bf{r},\bf{r}^\prime}S_{\bf{r}}^{x}S_{\bf{r'}}^{x}+J^y_{\bf{r},\bf{r}^\prime}S_{\bf{r}}^{y}S_{\bf{r'}}^{y}+J^{z}S_{\bf{r}}^{z}S_{\bf{r'}}^{z}]\\
&+{\bf D}_{\bf{r,r'}}\cdot (S_{\bf{r}}\times S_{\bf{r'}})+{{\bf h}\cdot {\bf S}_{\bf{r}}}.
     \end{split}
    \label{eq:spinmodel}
\end{equation}
This model incorporates Heisenberg exchange ($J^{x,y,z}$), Dzyaloshinskii-Moriya interaction (${\bf D}_{\bf{r,r'}}$), and an effective Zeeman field (${\bf h}$). 
Consistency is observed between the DMFT results and the predictions of this model. In regions of weak magnetic field ($m_z \ll t_0$) and suppressed SOC ($t_s \ll t_0$), the $J^{z}$-dominated Heisenberg interactions favor antiferromagnetic correlations between $A$ and $B$$C$ sublattices, stabilizing the $z$-FIM phase. Conversely, the Zeeman term promotes the STM phase. Increasing $t_s$ reverses the sign of Heisenberg  interactions via SOC, driving a transition to the SSM state~\cite{SM}.

\begin{figure}[!tb]
 \centering
\includegraphics[width=0.482\textwidth]{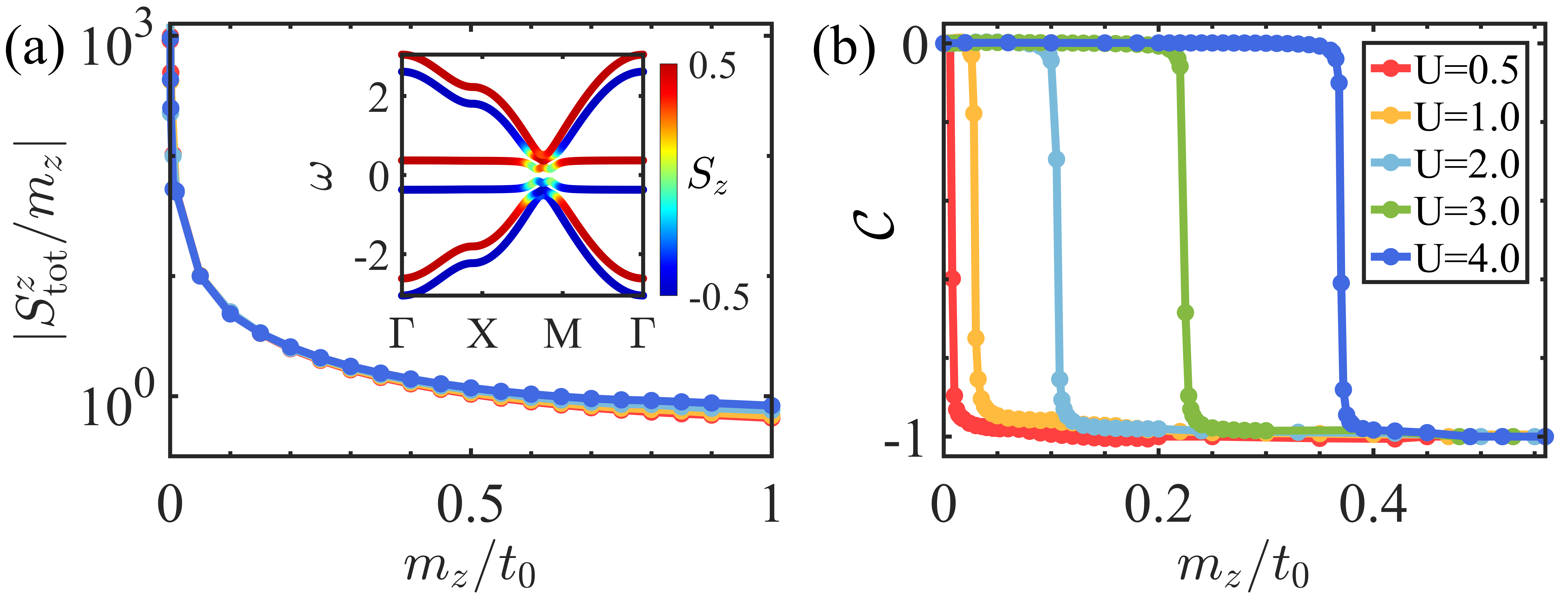} 
\caption{Intertwined emergence of spontaneous symmetry-breaking and topology. (a) Renormalized magnetization $S^z_{\rm tot}/m_z$ versus $m_z/t_{0}$ for varying $U$ and fixed $m_z=t_s$. Divergence at small $m_z$ signals spontaneous ferrimagnetic order. Inset: Spin texture along high-symmetry points for $U/t_0=1$, and $m_z=t_s=0.1t_0$ [$\mathcal{C}=-1$ from HF in (b)]. (b) Topological transitions along the $m_z=t_s$ line, where small $m_z$ and $t_s$ induce nonzero topological invariants for $U/t_0<1$.}
  \label{fig:mag_topo}
\end{figure}

{\it Intertwined emergence of magnetism and topology.} Having established the emergence of topological and magnetic orders, we now address the intriguing feature of the many-body ground state: the coexistence of magnetism and topology. Notably, two distinct coexistence regimes are observed. The first occurs under weak magnetic field ($m_z<t_0$) and suppressed SOC ($t_s<t_0$) where $z$-FIM order and topological states simultaneously emerge [Figs.~\ref{fig:topo_phases}(b) and~\ref{fig:mag_order}(a)]. The second manifests at strong SOC ($t_s\approx t_0$), exhibiting concurrent PM order and nontrivial topology [Figs.~\ref{fig:topo_phases}(c) and~\ref{fig:mag_order}(b)].

The intertwined ordered state emerging at weak $t_s$ and $m_z$ arises from both spontaneous symmetry breaking and topological order, despite explicit spin-rotational symmetry breaking by the magnetic field and SOC. As evidenced in Fig.~\ref{fig:mag_topo}(a), spontaneous symmetry breaking is confirmed by the divergence of renormalized magnetization $S^z_{\rm tot}/m_z$, where $S^z_{\rm tot}=1/N\sum_{\bf r} \langle S^z_{\bf r}\rangle$.
Simultaneously, small $t_s$ and $m_z$ perturbations generate nonzero topological invariants in the ground state for weak interactions $U/t_0<1$ [Fig.~\ref{fig:mag_topo}(b)]. This dual emergence originates from the Lieb lattice's flat band structure, where infinitesimal perturbations dramatically alter band properties~\cite{lieb1989two}, contrasting sharply with square lattices~\cite{wu2016realization,PhysRevResearch.5.L012006}. 
Consequently, weak interactions and SOC cooperatively induce magnetic spin polarization ($S^z_{\rm tot}\neq 0$) and topological spin winding ($\mathcal{C}=-1$) [inset of Fig.~\ref{fig:mag_topo}(a)]. We find that the topological ferrimagnetic phase is generic across the weak interaction regime, occupying extensive regions of the phase diagram for $m_z<t_0$ and $t_s<t_0$ (see Fig.~S3 in SM~\cite{SM}), indicating high experimental feasibility for observing this intertwined state.

\begin{figure}
\includegraphics[width=0.46\textwidth]{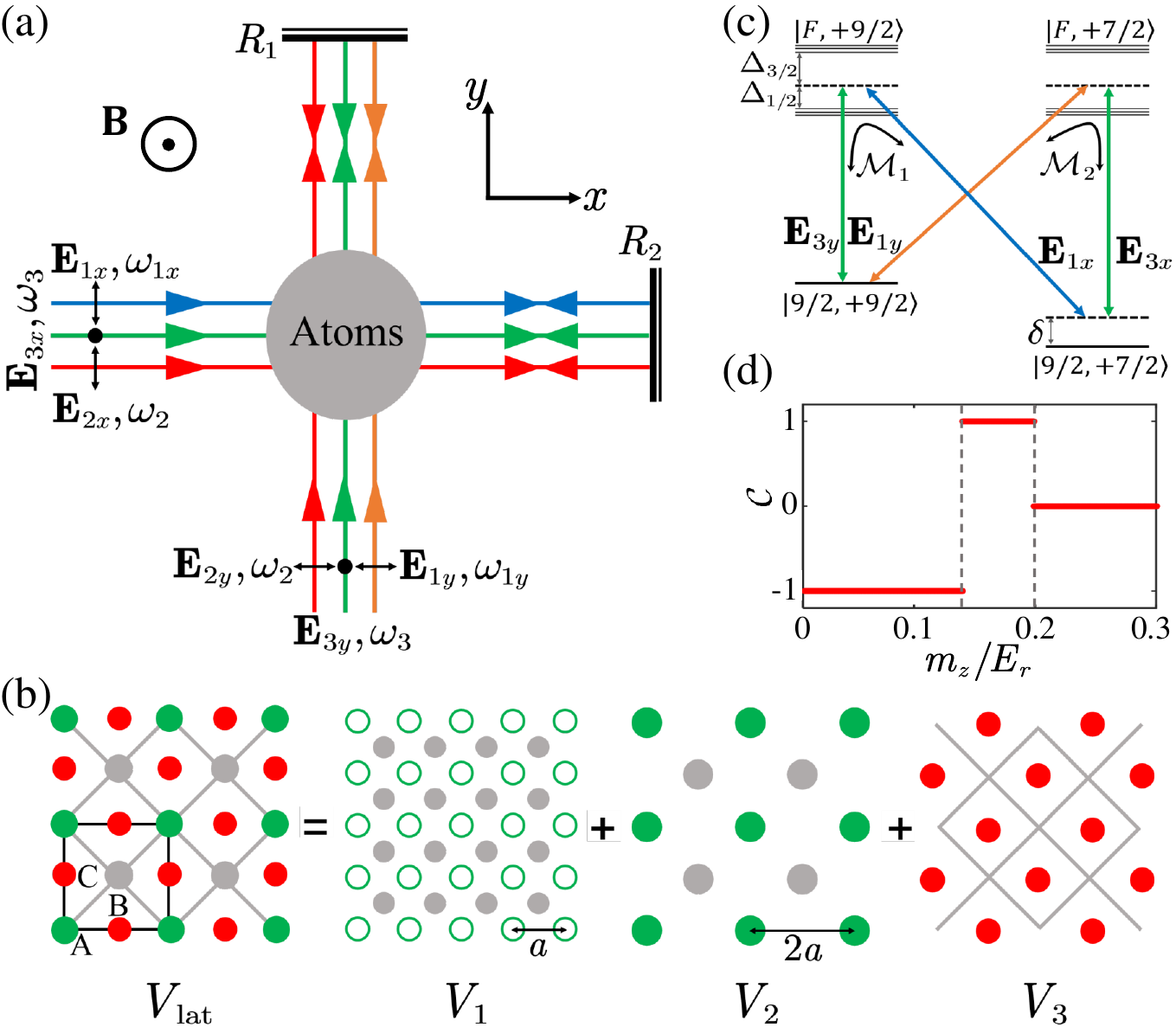}
	\caption{Experimental realization in ultracold atoms. (a) Three pairs of standing waves ${\bf E}_{nx}$ and ${\bf E}_{ny}$ propagating in the $x$- and $y$-directions simultaneously generate the Lieb lattice and Raman couplings, after reflection by mirrors $R_{1,2}$. Polarization orientations are marked by black dots and arrows. (b) Lieb lattice potential $V_{\rm lat}$ emerges from the superposition of three distinct lattices, where colored (gray) circles or lines mark potential minima (maxima). (c) Two Raman potentials $\mathcal{M}_{1,2}$ are induced via double-$\Lambda$ configurations formed by orthogonal polarization pairs $({\bf E}_{3y}, {\bf E}_{1x})$ and $({\bf E}_{1y}, {\bf E}_{3x})$. (d) Chern number ${\cal C}$ versus the Zeeman constant $m_z$. Here, we take $V_1=3E_{r}$, $V_2=-6E_{r}$, $V_3=-1.705E_{r}$, and $M_{01}=M_{02}=0.1E_{r}$, corresponding to $t_s/t_0\approx 0.09$.
	}\label{fig:realization}
\end{figure}
{\it Experimental proposal.} We propose a highly feasible experimental setup to realize a spin-orbit-coupled 2D Lieb lattice in ultracold atoms using optical Raman lattices~\cite{wu2016realization,song2019observation,PhysRevResearch.5.L012006}, as illustrated in Fig.~\ref{fig:realization}(a).
The total Hamiltonian reads
\begin{equation}\label{Ham_realization}
H=\frac{{\bf k}^2}{2m}+V_{\rm lat}({\bf r})+\mathcal{M}_1({\bf r}){\sigma}_x
+\mathcal{M}_2({\bf r}){\sigma}_y+m_z {\sigma}_z, 
\end{equation}
where ${\bf k}$ denotes the atomic momentum, $m$ is the atomic mass, $V_{\rm lat}({\bf r})$ represents the Lieb lattice potential, $\mathcal{M}_{1,2}({\bf r})$ correspond to Raman potentials that induce SOC, and $m_z=\delta/2$ with $\delta$ being the two-photon detuning. Below we outline the scheme, where details can be found in SM~\cite{SM}.

The Lieb lattice potential $V_{\rm lat}({\bf r})$ is formed by superimposing three optical lattices [Fig.~\ref{fig:realization}(b)]: 
(1) A square lattice ($V_1$) with spacing $a=\pi/k_0$, generated by orthogonal standing waves (${\bf E}_{1x}$ and ${\bf E}_{1y}$) of wavevector 
$k_0$. (2) A square lattice ($V_2$) with spacing $2a$, generated by beams ${\bf E}_{2x}$ and ${\bf E}_{2y}$ of $k_0/2$, 
phase-shifted by $\pm\pi/4$ to periodically eliminate $V_1$'s central sites with $2a\times2a$ supercell periodicity. (3) A
checkerboard lattice $V_3$, generated by coherent beams ${\bf E}_{3x}$ and ${\bf E}_{3y}$,
selectively enhancing potentials at $B$ and $C$ sites.
The combined potential landscape is given by~\cite{SM}
\begin{align}\label{V_latt}
V_{\rm lat}({\bf r})&=V_1\left[\cos^2(k_0 x)+\cos^2(k_0 y)\right]+\nonumber\\
&+V_2\left[\cos^2\left(\frac{k_0 x}{2}-\frac{\pi}{4}\right)+\cos^2\left(\frac{k_0 y}{2}+\frac{\pi}{4}\right)\right]\nonumber\\
&+V_3\left[\sin\left(k_0 x\right)+\sin\left(k_0 y\right)\right]^2,
\end{align}
where $V_{n} \propto E_{n}^2$, with $E_{nx}=E_{ny}=E_{n}$ and $E_{n\mu}$ 
denoting the field amplitude of the $n$-th beam propagating along the $\mu$-direction. 
For alkali-metal atoms, the Lieb lattice potential is spin-independent ($t_{\uparrow}=-t_{\downarrow}$). 
Raman couplings $\mathcal{M}_1({\bf r})=M_{01}\cos(k_0 x)\sin(k_0 y)$ and $\mathcal{M}_2({\bf r})=M_{02}\sin(k_0 x)\cos(k_0 y)$ (with $M_{01/02} \propto E_{3y/3x}E_{1x/1y}$~\cite{SM}) emerge from orthogonal polarization pairs $({\bf E}_{3y}, {\bf E}_{1x})$ and $({\bf E}_{1y}, {\bf E}_{3x})$, respectively, through a double-$\Lambda$ configuration with the condition $\omega_{1x}-\omega_{3}=\omega_{3}-\omega_{1y}$ [Fig.~\ref{fig:realization}(c)]. 
The relative antisymmetry between lattice and Raman potentials suppresses onsite spin flips, allowing only nearest-neighbor hopping $t_{\rm s}$. All parameters are independently tunable in experiments.

While the scheme is broadly applicable, we demonstrate it here using $\prescript{40}{}{\rm K}$ atoms, with $\ket{\uparrow}=\ket{F=9/2,m_F=+9/2}$ and $\ket{\downarrow}=\ket{9/2,+7/2}$ [Fig.~\ref{fig:realization}(c)]. 
Under typical parameters $V_1=3E_{r}$, $V_2=-6E_{r}$ and $V_3=-1.705E_{r}$ with $E_{r}\equiv k_{0}^{2}/(2m)$ being the recoil energy, 
the uniform coupling $M_{01}=M_{02}=0.1E_{r}$ leads to the spin-flip hopping $t_{\rm s}=0.09t_0$.
The noninteracting phase diagram obtained by diagonalizing Eq.~\eqref{Ham_realization} shows three characteristic regimes [Fig.~\ref{fig:realization}(d)], which is consistent with numerical calculations in Fig.~\ref{fig:topo_phases}(a). The presence of onsite interactions is expected to drive the spontaneous emergence of magnetic order [Fig.~\ref{fig:mag_topo}(a)].
Building upon the single-site resolution enabled by quantum gas microscopy~\cite{lebrat2024ferrimagnetism} and the tomographic reconstruction of nontrivial band topology in ultracold fermions~\cite{PhysRevResearch.5.L012006}, we anticipate that the predicted magnetic topological
phases in Lieb lattices could be experimentally observed in future studies using our proposed setup.

{\it Conclusion}. The interplay between many-body interactions and spin-orbit coupling drives the formation of exotic quantum phases in Lieb lattice systems, marked by the simultaneous emergence of topological invariants and spontaneous symmetry-breaking magnetic orders. Unlike previous approaches requiring strong interactions~\cite{PhysRevLett.116.225305,PhysRevLett.120.157205,PhysRevB.103.155108,PhysRevB.106.205107}, our scheme demonstrates the coexistence of topology and magnetic order even under weak interactions, highlighting its experimental feasibility with existing techniques. Future investigations could explore critical phenomena associated with topological phase transitions, emergent physics in engineered lattice geometries such as twisted bilayers hosting interaction-driven magnetism and fractional quantum Hall states~\cite{park2025ferromagnetism}, and the dynamical interplay between topological invariants and magnetic order in non-equilibrium many-body systems.

\textit{Acknowledgements.}
We acknowledge helpful discussions with Xibo Zhang, Jiaqi Wu, Hui Tan, Rui Cao, and Xiansi Wang. This work is supported by the National Natural Science Foundation of China (Grants No. 12374252, No. 12074431, No. 12274384 and No. 12204187), and the Science and Technology Innovation Program of Hunan Province under Grant No. 2024RC1046. L. Z. acknowledges support from the Innovation Program for Quantum Science and Technology (Grant No. 2021ZD0302000).

\bibliography{references.bib}

\clearpage
\onecolumngrid

\setcounter{figure}{0}
\setcounter{table}{0}
\setcounter{equation}{0}
\renewcommand{\thefigure}{S\arabic{figure}}
\renewcommand{\thetable}{S\arabic{table}}
\renewcommand{\theequation}{S\arabic{equation}}

\centerline{\Large{\bf Supplemental Materials}}

\maketitle

\renewcommand{\theenumi}{\Roman{enumi}}  
\renewcommand{\theenumii}{\Alph{enumii}} 
\renewcommand{\theenumiii}{\arabic{enumiii}}

\section*{CONTENTS}
\addcontentsline{toc}{chapter}{Table of Contents}
\begin{enumerate}
 \item \hyperref[chapter:DMFT]{Dynamical Mean-Field Theory\hfill 8}
 \item \hyperref[chapter:Hartree-Fock]{Hartree-Fock method in the weakly interacting regime\hfill 10}
  \item \hyperref[chapter:eff spin model]{Effective spin model in the strongly interacting regime\hfill 11}
  \item \hyperref[chapter:topo invariant]{Topological invariant and spectral function\hfill 12} 
  \begin{enumerate}
      \item \hyperref[section:Chern number]{Chern number\hfill 12}
      \item \hyperref[section:Bott index]{Bott index\hfill 13}
      \item \hyperref[section:Spectral function]{Spectral function\hfill 14}
    \end{enumerate}
  
  \item \hyperref[chapter:mag invariant]{Magnetic order parameter\hfill 14} 
  \item \hyperref[chapter:self energy]{Self energy analysis for the topology of interacting systems\hfill 15} 
  \item \hyperref[chapter:Experimental realization]{Experimental realization\hfill 15} 
  \begin{enumerate}
      \item \hyperref[section:Lattice and Raman potentials]{Lattice and Raman potentials\hfill 16}
      \item \hyperref[section:Tight-binding model]{Tight-binding model\hfill 18}
    \end{enumerate}
\end{enumerate}

\vspace{1em}

\section{Dynamical Mean-Field Theory}\label{chapter:DMFT}
Dynamical mean-field theory (DMFT) is a powerful theoretical framework that is exact in infinite dimensions and serves as a good approximation for finite-dimensional systems. It has been extensively studied and applied to a wide range of strongly correlated systems. The central idea of DMFT is to map the many-body lattice problem onto a single-site impurity model coupled to a non-interacting fermionic bath. This mapping allows for the self-consistent solution of the impurity problem, capturing local quantum fluctuations while treating long-range correlations as a Weiss mean field. Base on this approximation, we transform the problem of solving the initial quantum many-body Hubbard model into the problem of solving the effective action of the impurity. The local effective action for the impurity site is given by
\begin{equation}
	\begin{split}
		\mathrm{S}_\mathrm{eff}^{(\mathbf{0})}=&\int_{0}^{\beta}d\tau d\tau^{\prime}\sum_{\nu,\nu^{\prime}}\begin{pmatrix}c_{\nu,\mathbf{0}}^{\star}(\tau)\\c_{\nu,\mathbf{0}}(\tau)\end{pmatrix}^{T}\mathcal{G}_{\mathbf{0}}^{-1}(\tau-\tau^{\prime})\begin{pmatrix}c_{\nu^{\prime},\mathbf{0}}(\tau^{\prime})\\c_{\nu^{\prime},\mathbf{0}}^{\star}(\tau^{\prime})\end{pmatrix}+Un_{\uparrow,\mathbf{0}}n_{\downarrow,\mathbf{0}},
	\end{split}
	\label{eq:Seff}
\end{equation}
where $\mathcal{G}_{\mathbf{0}}^{-1}(\tau-\tau^{\prime})$ represents a local non-interacting propagator, which acts as a dynamical Weiss mean field. It effectively simulates the influence of all other sites in the system. $\mathrm{S}_{\rm eff}^{(0)}$ enables the calculation of all local correlation functions for the original Hubbard model.

In practice, directly solving the effective action of the impurity can be computationally challenging. Therefore, we further map the impurity problem to an Anderson impurity model, which is more tractable for numerical calculations. For each impurity site, the effective Anderson impurity Hamiltonian can be written as
\begin{equation}
	\begin{split}
		\mathrm{H}_\mathrm{imp}^{(r)}=U\hat{n}_{\uparrow}\hat{n}_{\downarrow}-\sum_{\nu}\mu_{\nu}\hat{n}_{\nu}+\sum_{l,\nu}\epsilon_{l,\nu}^{(r)}\hat{a}_{l,\nu}^{\dagger}\hat{a}_{l,\nu}+\sum_{l,\nu}\left(V_{l,\nu}^{(r)}\hat{a}_{l,\nu}^{\dagger}\hat{c}_{\nu}+W_{l,\nu}^{(r)}\hat{a}_{l,\bar{\nu}}^{\dagger}\hat{c}_{\nu}+\mathrm{H.c.}\right).
	\end{split}
	\label{eq:Seff}
\end{equation}
Here, $r$ denotes the impurity site, $l$ labels the bath orbitals, and $\nu$ represents the spin state. The operator $\hat{a}_{l,\nu}$ describes the non-interacting fermions in the bath, with $\epsilon_{l,\nu}$ being their corresponding energies. The operator $\hat{c}_{l,\nu}$ describes the local impurity fermions. The coupling between the bath and the impurity site is characterized by the parameters $V_{l,\nu}$ and $W_{l,\nu}$, which represent the spin-conserving and spin-flipping interactions, respectively.

In our work, we primarily use numerical methods such as exact diagonalization~\cite{PhysRevB.40.7406} and Lanczos diagonalization~\cite{PhysRevB.67.161103,PhysRevB.86.075141} to solve the impurity problem. These methods allow us to obtain the impurity Green's function $G(i\omega_n)$. Once the impurity Green's function is computed, the local self-energy $\Sigma(i\omega_n)$ for each site can be extracted using the Dyson equation
\begin{equation}
	\begin{split}
		\Sigma(i\omega_n)=\mathcal{G}^{-1}(i\omega_n)-G^{-1}(i\omega_n),
	\end{split}
	\label{eq:Dyson}
\end{equation}
where $\mathcal{G}^{-1}(i\omega_n)$ is the Weiss Green's function, which represents the effective bath coupled to the impurity site. $\omega_n$ is the Matsubara frequency.

In the framework of DMFT, we assume that the impurity self-energy $\Sigma(i\omega_n)$ is identical to the lattice self-energy $\Sigma_{\rm lat}(i\omega_n)$. This assumption allows us to employ the Dyson equation in the real-space representation to compute the interacting lattice Green's function
\begin{equation}
	\begin{split}
		\mathbf{G}_\mathrm{lat}^{-1}(i\omega_{n})=\mathbf{G}^{-1}(i\omega_{n})-\mathbf{\Sigma}_\mathrm{lat}(i\omega_{n}).
	\end{split}
	\label{eq:Glatt}
\end{equation}
The self-consistency loop in DMFT is closed by using the Dyson equation to obtain a new local non-interacting propagator. Subsequently, new Anderson impurity parameters are updated by minimizing the difference between the old and new propagators. This procedure is iterated until convergence is achieved.

\begin{figure}[H]
	\centering
\includegraphics[width=0.65\textwidth]{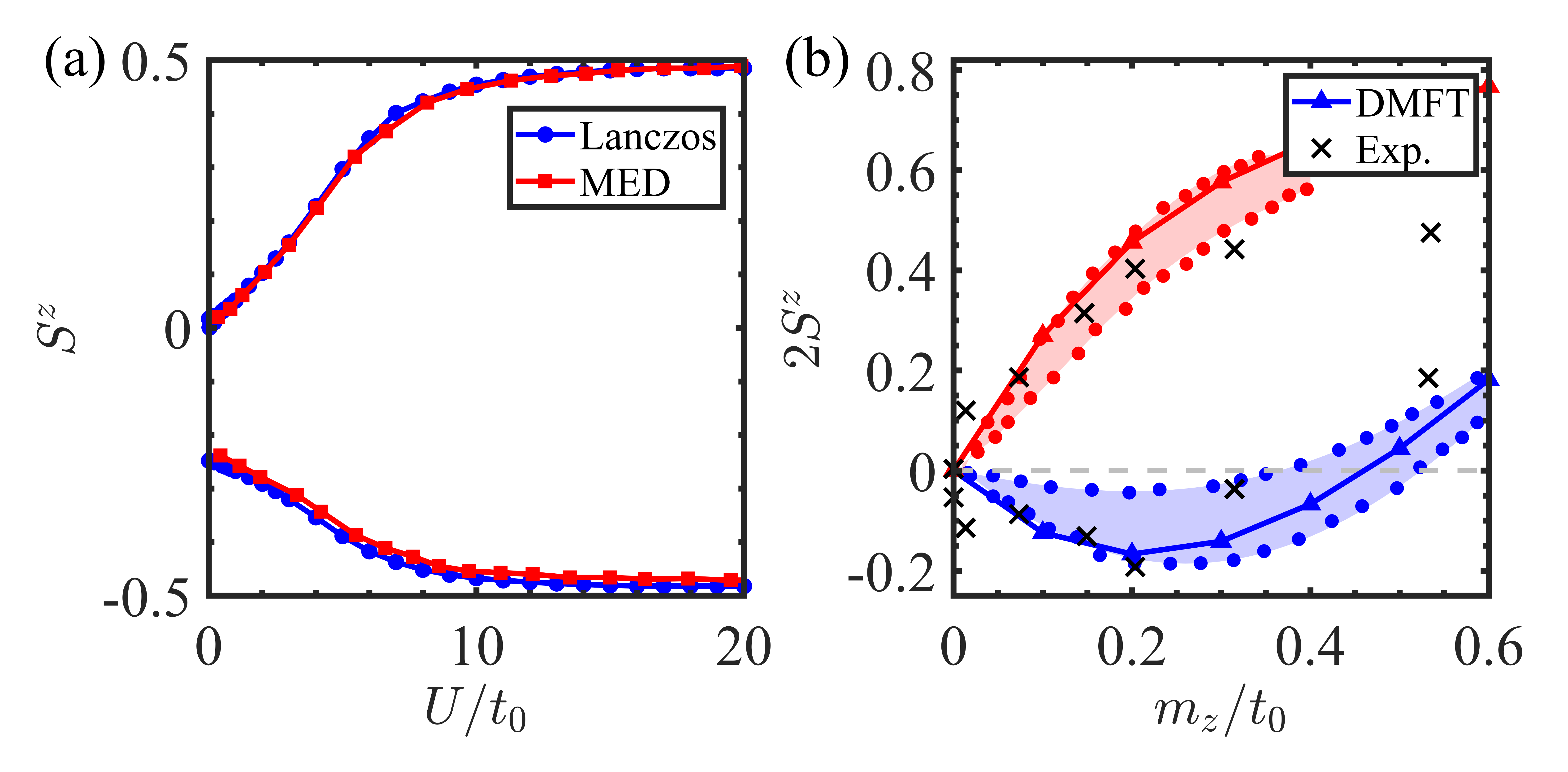} 
	\caption{Magnetic order parameters without SOC in two-dimensional Lieb lattices. (a) Magnetic order $S^z$ as a function of $U/t_0$ for temperature $T=0.05t_0$, and magnetic field $m_z=0$. Blue circles represent our DMFT results (Lanczos solver), while red squares show the modified exact diagonalization data (MED) from Ref.~\cite{nguyen2016dynamical}. (b) Magnetic order in the presence of symmetry-breaking magnetic field $m_z$ at $T=0.4t_0$, and  $U=6t_0$, which is consistent with experimental observations and quantum Monte Carlo simulations in Ref.~\cite{lebrat2024ferrimagnetism}. Triangles mark our DMFT result, shaded areas represent numerical data from determinant quantum Monte Carlo simulations, and black crosses denote experimental data.
	}
	\label{fig:ferri}
\end{figure}

To validate the DMFT approach, we compute magnetic ordering of the two-dimensional Lieb lattice in absence of spin-orbit coupling (SOC). The results were compared with existing theoretical and experimental results~\cite{nguyen2016dynamical,lebrat2024ferrimagnetism}, as shown in Fig.~\ref{fig:ferri}. Fig.~\ref{fig:ferri}(a) demonstrates the emergence of ferrimagnetic order even at relatively small value $U$, consistent with previous theoretical work~\cite{nguyen2016dynamical}. In Fig.~\ref{fig:ferri}(b), we present magnetic order in the presence of a finite symmetry-breaking magnetic field. Comparison has been made with those from experiments (black cross) and from quantum Monte-Carlo simulations (shaded region)~\cite{lebrat2024ferrimagnetism}. The  agreements with both experimental data and quantum Monte-Carlo simulations  demonstrate the accuracy of our DMFT implementation, and its ability to connect theory with experiment.

\begin{figure}[H]
	\centering	\includegraphics[width=0.90\textwidth]{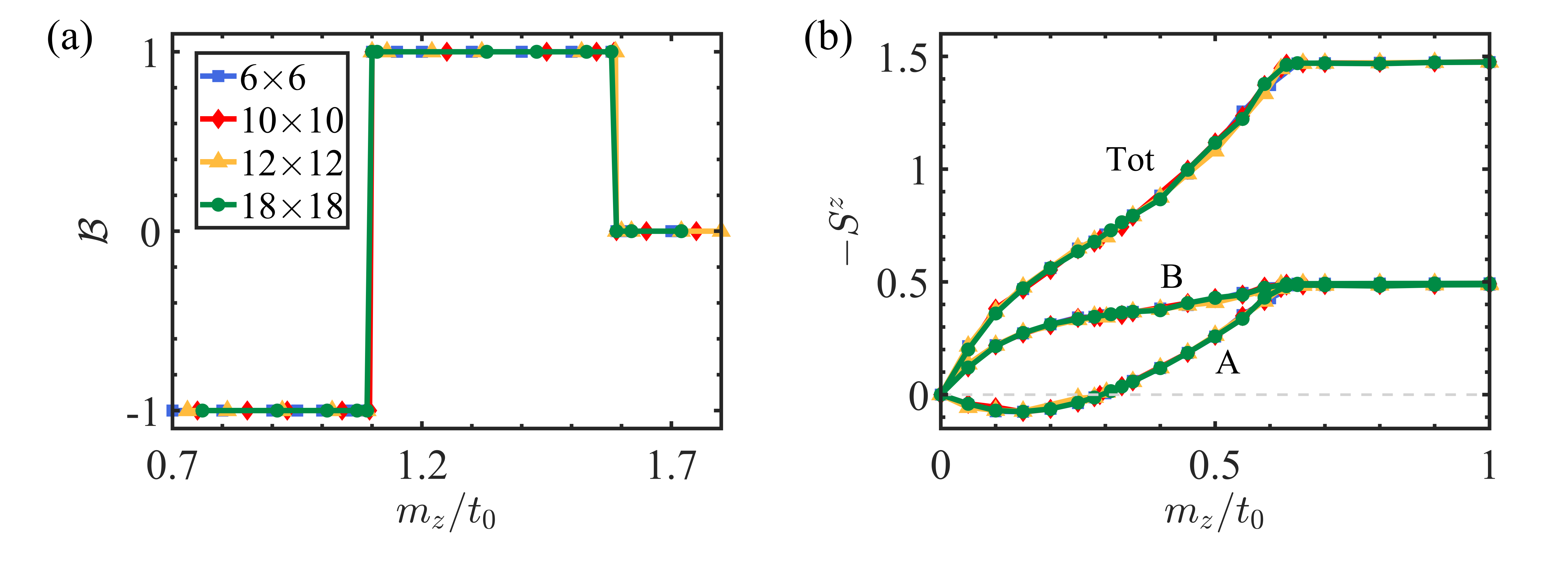} 
	\caption{Topological invariant and magnetic order for different system sizes calculated using DMFT. (a) Bott index as a function of $m_z$ for $U=3t_{0}$ and $t_{s}=0.5t_{0}$. (b) Spin order parameter $\mathbf{\langle S}^z\rangle$ as a function of $m_z$ for $U=8t_{0}$ and $t_s=0.5t_{0}$. The results are shown for different system sizes of $3\times 6 \times 6$ (blue squares), $3\times10 \times 10$ (red diamonds), $3\times12 \times 12$ (yellow triangles), and $3\times18 \times 18$ (green circles).
	}
	\label{fig:size_check}
\end{figure}

Subsequently, we incorporate SOC into the Lieb system to investigate its magnetic order and topological properties. In our DMFT calculations, the bath orbital is truncated to be at maximum orbital number $n_{\rm bath}= 5$, and the temperature is set to be $T=0.02t_0$. To assess the finite-size effects, we perform calculations for different cell sizes, and find that both the topological and magnetic phase boundaries converge even for a $3\times6 \times 6$ lattice size, as shown in Fig.~\ref{fig:size_check}. In our calculations, we primarily employ a $3\times12\times12$ lattice size to obtain the topological and magnetic phase diagrams.

\section{Hartree-Fock method in the weakly interacting regime}\label{chapter:Hartree-Fock}
In the weak interaction regime ($U\ll t$), Hartree-Fock method is a good approximation, and can be used to obtain band structures and magnetic order~\cite{das2024realizing}. Within Hartree-Fock approximation, the interaction term of Eq. (1) in the main text is decoupled as~\cite{PhysRevB.104.235115} 
\begin{equation}
	U n_{c_{\boldsymbol{r},\uparrow}} n_{c_{\boldsymbol{r},\downarrow}} = U \langle n_{c_{\boldsymbol{r},\uparrow}} \rangle n_{c_{\boldsymbol{r},\downarrow}} + U n_{c_{\boldsymbol{r},\uparrow}} \langle n_{c_{\boldsymbol{r},\downarrow}} \rangle - U \langle n_{c_{\boldsymbol{r},\uparrow}} \rangle \langle n_{c_{\boldsymbol{r},\downarrow}} \rangle,
	\label{eq:HF}
\end{equation}
where $\boldsymbol{r}$ denotes the site index. Note here that the last term on the right hand is actually the energy shift, which does not change the self-consistency loop. Based on Eq.~(\ref{eq:HF}), the two-body interaction is decoupled into the single-particle representation. After introducing a set of momentum-space basis $\psi_{\boldsymbol{k}} = \{ c_{\boldsymbol{k},A,\uparrow}, c_{\boldsymbol{k},B,\uparrow}, c_{\boldsymbol{k},C,\uparrow}, c_{\boldsymbol{k},A,\downarrow}, c_{\boldsymbol{k},B,\downarrow}, c_{\boldsymbol{k},B,\downarrow} \}^T$, Eq. (1) in the main text can be written as $H=\sum_{\bf k}\psi_{\boldsymbol{k}}^\dagger \mathcal{H}_{\boldsymbol{k}} \psi_{\boldsymbol{k}}$ by taking Fourier transformation, where $\mathcal{H}_{\boldsymbol{k}}$ is given by
\begin{equation}
	\mathcal{H}_{\boldsymbol{k}}= \left[ 
	\begin{smallmatrix}
		-\mu +m_z + U \langle n_{A,\downarrow} \rangle & -t_\uparrow & -t_\uparrow & 0 & t_s & t_s \\
		-t_\uparrow & -\mu +m_z + U \langle n_{B,\downarrow} \rangle & 0 & t_s & 0 & 0 \\ 
		-t_\uparrow & 0 & -\mu +m_z + U \langle n_{C,\downarrow} \rangle & t_s & 0 & 0 \\
		0 & t_s & t_s & -\mu -m_z + U \langle n_{A,\uparrow} \rangle & t_\downarrow & t_\downarrow \\
		t_s & 0 & 0 & t_\downarrow & -\mu -m_z + U \langle n_{B,\uparrow} \rangle  & 0 \\
		t_s & 0 & 0 & t_\downarrow & 0 & -\mu -m_z + U \langle n_{C,\uparrow} \rangle
	\end{smallmatrix}
	\label{eq:HFA}
	\right].
\end{equation}
Here, the chemical potential $\mu$ is introduced to adjust the total density in the unit cell, guaranteeing the half-filling condition with 
\begin{equation}
	\begin{split}
		n &= \frac{1}{3N_{\boldsymbol{k}}} \sum_{\boldsymbol{k},\alpha,\nu} \langle n_{\boldsymbol{k}, \alpha,\nu} \rangle_{\mathrm{HF}} \\
		& = \frac{1}{3N_{\boldsymbol{k}}} \sum_{\boldsymbol{k}, i} f(\varepsilon_{\boldsymbol{k},i} - \mu),
	\end{split}
	\label{eq:find_mu}
\end{equation}
where $\alpha = A, B, C$ is the site index in the unit cell, $i$ is the Hartree-Fock band index, $N_{\boldsymbol{k}}$ is the total mesh number in the first Brillouin zone, and $f$ is the Fermi-Dirac distribution.

\begin{figure}[htbp]
	\centering
	\includegraphics[width=0.645\textwidth]{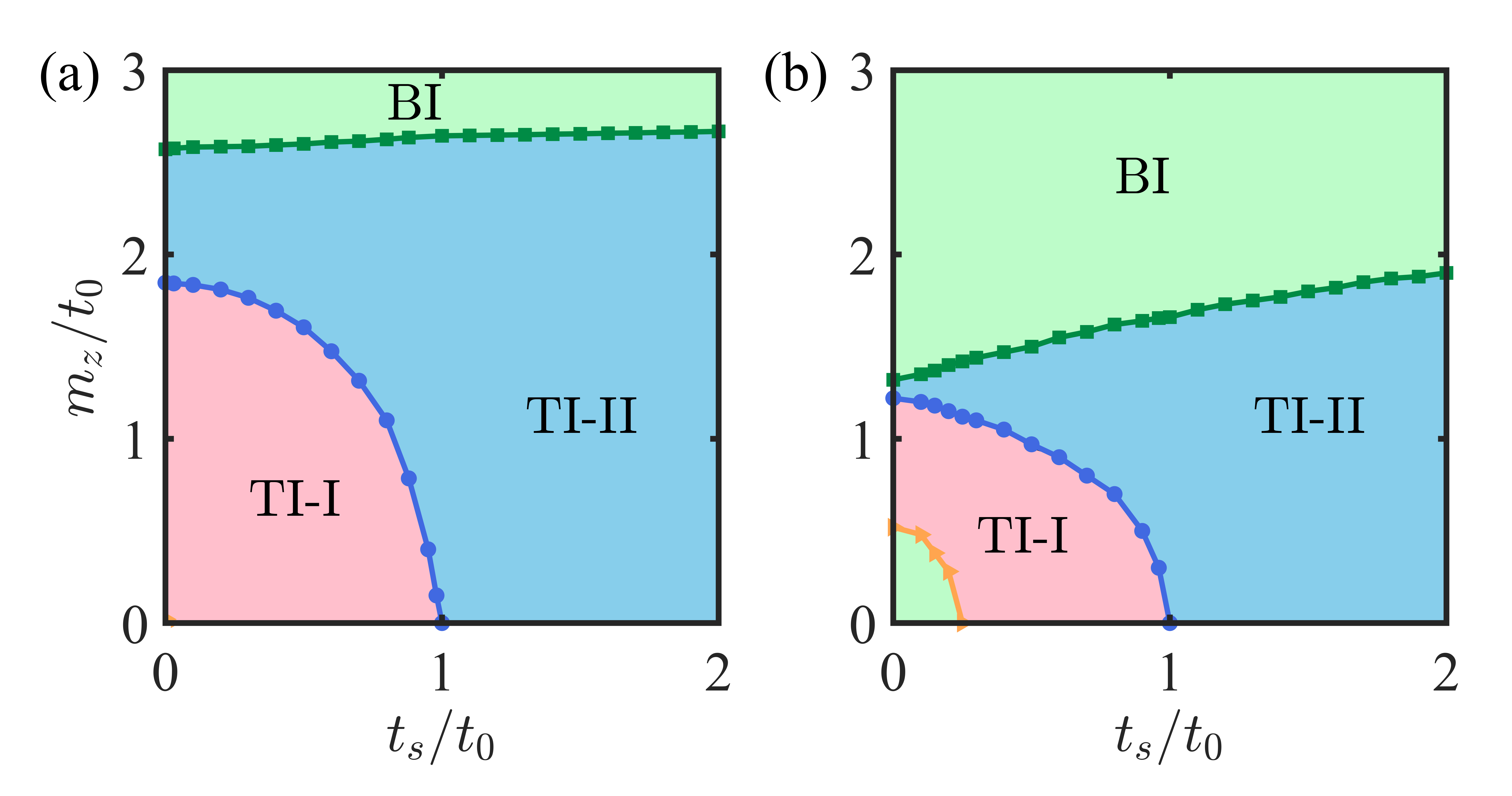} 
	\caption{Topological phase diagrams of interacting fermions in the two-dimensional Raman Lieb lattice at half-filling for the interaction strengths (a) $U=0.5t_{0}$ and (b) $U=3t_0$, obtained by Hartree-Fock approximation. The system supports three quantum phases, including a trivial (BI) and two nontrivial (TI-I and TI-II) phases. 
	}
	\label{fig:HF_U0.5}
\end{figure}

Eq.~(\ref{eq:HFA}) can be solved self-consistently for a fixed filling. During the self-consistency loop, we change the chemical potential and update the particle density in each step, until the density difference for all site meets the convergence condition $|n_\alpha(\mathrm{new}) - n_\alpha(\mathrm{old}) | < 1.0\times 10^{-6}$.
Using this approach, we obtain topological phase diagrams for the interaction strengths $U=0.5t_{0}$ and $3t_{0}$, as illustrated in Fig.~\ref{fig:HF_U0.5}. We observe three topologically distinct phases, including one trivial and two nontrivial phases. We find that the phase diagram is robust against onsite interactions, compared to Fig. 2(a) in the main text. In addition, the Hartree-Fock method provides accurate results in the weak interaction regime, and yields excellent agreement with the prediction of DMFT, as shown in Fig.~\ref{fig:HF_U0.5}(b). Note here that, we choose $N_{\boldsymbol{k}}=300\times300$ to determine the Chern number $\mathcal{C}$.

\section{Effective spin model in the strongly interacting regime}\label{chapter:eff spin model}
In the strongly correlated limit $(U\gg t)$, double occupancy in each lattice site is strongly suppressed. To describe this case, we derive an effective spin-exchange model by introducing the projection operator $\mathcal{P}$, which restricts the system to the states where each site has exactly one particle, and $\mathcal{Q}=1-\mathcal{P}$ including all states with at least one site having double or higher occupancy. The Hamiltonian consists of $H_t$ and $H_U$, where $H_t$ describes the hopping between different lattice sites, and $H_U$ represents the onsite interaction. By treating the hopping terms as perturbations, an effective  model in the deep Mott-insulating regime is derived 
\begin{equation}  \mathcal{H}_\mathrm{eff}=\mathcal{P}H_{t}\mathcal{Q}\frac{1}{E-\mathcal{Q}H_{U}\mathcal{Q}-\mathcal{Q}H_{t}\mathcal{Q}}\mathcal{Q}H_{t}\mathcal{P},
	\label{eq:H_eff}  
\end{equation}
where $E$ is the energy of the system in the $\mathcal{P}$ subspace. By keeping terms up to second order in perturbation theory, Eq.~(\ref{eq:H_eff}) is reduced to~\cite{PhysRevLett.111.205302,PhysRevA.99.063603} 
\begin{equation}  
\mathcal{H}_\mathrm{eff}=\mathcal{P}H_{t}\mathcal{Q}\frac{1}{E-\mathcal{Q}H_{U}\mathcal{Q}}\mathcal{Q}H_{t}\mathcal{P}
	\label{eq:second_order_H}.  
\end{equation}

We first consider a two-site system at half filling, where the states under $\mathcal{P}$ are
\begin{equation}  \begin{split}|\uparrow,\uparrow\rangle,|\uparrow,\downarrow\rangle,|\downarrow,\uparrow\rangle,|\downarrow,\downarrow\rangle, 
	\end{split}
	\label{eq:PQ}  
\end{equation}
and those under $\mathcal{Q}$ are
\begin{equation}  \begin{split}|\uparrow\downarrow,0\rangle,|0,\uparrow\downarrow\rangle.
	\end{split}
	\label{eq:PQ}  
\end{equation}
Then, we calculate all the matrix elements required for Eq.~(\ref{eq:second_order_H}), where eigenenergies of the ground and excited states are given in Table.~\ref{tab:E}. By extending the two-site model to the whole lattice, we finally obtain a second-order spin-exchange Hamiltonian at half filling

\begin{equation}
\begin{split}
H_{\mathrm{eff}}=
&\sum_{\left\langle\mathrm{r,r'}\right\rangle}[J^x_{\bf{r},\bf{r}^\prime}S_{\bf{r}}^{x}S_{\bf{r'}}^{x}+J^y_{\bf{r},\bf{r}^\prime}S_{\bf{r}}^{y}S_{\bf{r'}}^{y}+J^{z}S_{\bf{r}}^{z}S_{\bf{r'}}^{z}]+{\bf D}_{\bf{r,r'}}\cdot (S_{\bf{r}}\times S_{\bf{r'}})+{{\bf h}\cdot {\bf S}_{\bf{r}}},
     \end{split}
    \label{eq:2}
\end{equation}
where $\mathbf{S}_r\equiv c^\dagger_{ r,\nu}\sigma_{\nu\nu^\prime}c_{{ r},\nu^\prime}$ for site $r$, with $\sigma_{\nu\nu^\prime}$ denoting Pauli matrices. Here, the Heisenberg coupling $J$, Dzyaloshinskii-Moriya term $\mathbf{D}$, and effective magnetic field $\mathbf{h}$ are presented in Table.~\ref{tab:spinpara}. Depending on the values of $t_s$ and $m_z$, the coupling parameters in Eq.~(\ref{eq:2}) compete with each other. In our case, we take $t_\uparrow=-t_\downarrow=t_0$, and find that the influence of spin-orbit coupling is taken into account through modifying the values of $J_{x,y,z}$, since ${\bf D}$ always vanishes. This effective spin-exchange model provides an underlying picture for the magnetic phase diagram in the strongly interacting regime, such as Fig. 3(b) in the main text with $U=8t_{0}$. In the  regime with $t_s/t_{0}\ll 1$ and $m_z/t_{0}\ll 1$, the $J_z$ term dominates, and leads to the ferrimagnetic phase. In contrast, for larger spin-orbit coupling with $t_s/t_{0}\gg 1$, the sign of effective Heisenberg couplings is changed, which induces a spin-spiral phase. For larger magnetic field with $m_z/t_{0} \gg 1$, the spin tends to align along the magnetic field with partially and saturated magnetic phases appearing instead.

\begin{table}[htbp]
	\centering
	\caption{Energy levels of the ground and excited states for a two-site problem in the strongly interacting limit.}
	\resizebox{5cm}{!}{
		\fontsize{6}{16}\selectfont
		\begin{tabular}{c@{\hspace{1cm}}c}
			\hline
			\(\mathbf{state}\) & \(\mathbf{Energy}\)\\
			\hline
			\(|\left \downarrow;\downarrow \right\rangle\) & \(2\mu_{\downarrow}-2m_z\)\\
			\(|\left \downarrow;\uparrow \right\rangle\) & \(\mu_{\uparrow}+\mu_{\downarrow}\)\\
			\(|\left \uparrow;\downarrow \right\rangle\) & \(\mu_{\uparrow}+\mu_{\downarrow}\)\\
			\(|\left \uparrow;\uparrow \right\rangle\) & \(2\mu_{\uparrow}+2m_z\)\\
			\(|\left \uparrow\downarrow;0\right\rangle\) & \(U_{\uparrow\downarrow}+\mu_{\uparrow}+\mu_{\downarrow}\)\\
			\(\left|0;\uparrow\downarrow\right\rangle\) & \(U_{\uparrow\downarrow}+\mu_{\uparrow}+\mu_{\downarrow}\)\\
			\hline
		\end{tabular}
	}
	\label{tab:E}
\end{table}

\begin{table}[H]
	\centering
	\caption{Summary of interaction parameters for the effective spin-exchange model.}
	\resizebox{6cm}{!}{
		\fontsize{6}{16}\selectfont
		\begin{tabular}{c@{\hspace{1cm}}c}
			\hline
			\(J_{AB}^x, J_{AC}^y \) & \(\frac{4(t_\uparrow t_\downarrow-(t_s)^2)}{U} \)\\
			\(J_{AB}^y, J_{AC}^x\) & \(\frac{4(t_\uparrow t_\downarrow+(t_s)^2)}{U}\)\\
			\(J^z\) & \(\frac{2(t_\uparrow^{2}+t_\downarrow^{2})}{U}-\frac{4U(t_s)^2}{U^2-4m_{z}^2}\)\\
			\(D_{rr'}\) & \(\frac{4U(t_\uparrow t_s^\mathrm{0}+t_\downarrow t_s)}{U^2-m_{z}^2}(\mathbf{r}-\mathbf{r'})\)\\
			\(h_{x}\) & \(\frac{2m_{z}(t_\uparrow t_s+t_\downarrow t_s)}{U^2-m_{z}^2}\)\\
			\(h_{y}\) & \(-\frac{2m_{z}(t_\uparrow t_s+t_\downarrow t_s)}{U^2-m_{z}^2}\)\\
			\(h_{z}\) & \(-\frac{4m_{z}(t_s)^2}{U^2-4m_{z}^2}+2m_z\)\\
			\hline
		\end{tabular}
	}
	\label{tab:spinpara}
\end{table}

\section{Topological Invariant and spectral function}\label{chapter:topo invariant}
Topological properties of the two-dimensional system can be characterized by two kinds of topological invariants, such as Chern number $\mathcal{C}$ and Bott index $\mathcal{B}$. While $\mathcal{C}$ elucidates $k$-space Berry flux distribution, $\mathcal{B}$ directly detects edge states through open-boundary density matrix truncation. Here, 
both topological invariants are utilized to identify interaction-introduced topology. In addition, we employ the spectral function to further characterize topological properties of the interacting system, which provides insight into the system’s edge states.

\subsection{Chern number}\label{section:Chern number}
The Chern number, a well-established topological invariant in $k$-space, is calculated from the Berry curvature in reciprocal space~\cite{PhysRevB.87.174402,PhysRevApplied.9.024029,PhysRevB.87.174427},
\begin{equation}
	\mathcal{C}_n = \frac{1}{2\pi} \int_{\mathrm{B.Z.}}\mathcal{F}_{xy}(\mathbf{k})d^{2}k,
	\label{eq:chern}
\end{equation}
where $\mathcal{F}_{xy}(\mathbf{k})$ is the gauge-invariant Berry curvature for the $n$-th band, with
\begin{equation}
	\mathcal{F}_{xy}(\mathbf{k})=i\,\mathrm{Tr}\left[P_n\left(\partial_{k_x}P_n\partial_{k_y}P_n-\partial_{k_y}P_n\partial_{k_x}P_n\right)\right],
	\label{eq:Fxy}
\end{equation}
and $P_n=|\psi_n\rangle\langle\psi_n|$ is the spectral projector for the $n$-th Hartree-Fock wavefunctions $|\psi_n\rangle$. The integral is taken over the first Brillouin zone. In this work, we focus on topological properties of the many-body ground states, which are characterized by the total Chern number, $\mathcal{C} \equiv \sum_{n\in\mathrm{occ.}} \mathcal{C}_n$, by summing over the Chern numbers of the occupied bands.

\subsection{Bott index}\label{section:Bott index}
The Bott index is a real-space topological invariant that serves as a powerful tool for characterizing topological phases for fermionic systems. Unlike momentum-space topological invariants, such as the Chern number, the Bott index can be directly computed from the real-space eigenstates of a fermionic system, making it particularly useful for studying disordered or finite systems. In this section, we provide a detailed explanation of the Bott index.

The fermionic band structure is obtained by solving the eigenvalue problem of the topological Hamiltonian, which exhibits the same topological properties as the original Hamiltonian in a gapped system.
\begin{equation}
	\eta H_{\mathrm{topo}}\Psi=\Psi\eta E
	\label{eq:band}
\end{equation}
where $E$ is the diagonal matrix
of eigenvalues, $\Psi$ is the matrix eigenstates of the Hamiltonian, and $\eta=\mathbb{1}$ for fermionic system.
The topological Hamiltonian $H_{\mathrm{topo}}$ is defined as~\cite{Wang_2013}
\begin{equation}  H_{\mathrm{topo}}=H_0+\Sigma(\omega\rightarrow0),
\label{Bott_selfenergy}
\end{equation}
Here, $H_0$ is the non-interacting Hamiltonian, and $\Sigma(\omega\rightarrow0)$ is the local self-energy of the interacting systems. In our simulations, the self-energy $\Sigma(\omega)$ is obtained by DMFT, providing a powerful tool for studying topology of the strongly interacting systems.

After obtaining the self-energy of the interacting system, $H_{\mathrm{topo}}$ is constructed, whose eigenstates can be utilized to define the Bott index. For a given set of states $\mathcal{N}$, a projection operator $P_{\mathcal{N}}$ onto these states is given by
\begin{equation}
	P_{\mathcal{N}}=\Psi\Gamma_{\mathcal{N}}\Psi^{\dagger}
	\label{eq:P}
\end{equation}
\begin{equation}
	\Gamma_{\mathcal{N}}=\begin{pmatrix}0&0\\0&\mathbb{1}\end{pmatrix},
	\label{eq:Pp}
\end{equation}
where $\Gamma_{\mathcal{N}}$ is a diagonal matrix with entries 1 for states in $\mathcal{N}$ and 0 otherwise. The rescaled coordinates $X$ and $Y$ are defined as
\begin{equation}
	\begin{split}
		X=i_x/N_x,\quad
		Y=i_y/N_y\in[0,1),
	\end{split}
	\label{eq:P1}
\end{equation}
where $i_x$ and $i_y$ are spatial indices of the unit cells, and $N_x$ and $N_y$ are the number of cells in the $x$ and $y$ directions, respectively.

The Bott index $\mathcal{B}_{\mathcal{N}}$ is then constructed from the projected position operators $P_{\mathcal{N}}e^{i\pi X}P_{\mathcal{N}}$ and $P_{\mathcal{N}}e^{i\pi Y}P_{\mathcal{N}}$. These operators can be expressed in terms of the eigenstates $\Psi$ and the projection matrix $\Gamma_\mathcal{N}$
\begin{equation}
	\begin{split}
		P_{\mathcal{N}}e^{i\pi X}P_{\mathcal{N}}=\Psi\Gamma_{\mathcal{N}}\Psi^{\dagger}e^{i\pi X}\Psi\Gamma_{\mathcal{N}}\Psi^{\dagger}\\
		P_{\mathcal{N}}e^{i\pi Y}P_{\mathcal{N}}=\Psi\Gamma_{\mathcal{N}}\Psi^{\dagger}e^{i\pi Y}\Psi\Gamma_{\mathcal{N}}\Psi^{\dagger}.
	\end{split}
	\label{eq:P2}
\end{equation}
Here the matrices $U$ and $V$ is defined as
\begin{equation}
	\begin{split}
		&\Gamma_{\mathcal{N}}\Psi^\dagger e^{i\pi X}\Psi\Gamma_{\mathcal{N}}\equiv\begin{pmatrix}0&0\\0&U\end{pmatrix}\\
		&\Gamma_{\mathcal{N}}\Psi^{\dagger}e^{i\pi Y}\Psi\Gamma_{\mathcal{N}}\equiv\begin{pmatrix}0&0\\0&V\end{pmatrix}.
	\end{split}
	\label{eq:P3}
\end{equation}
Then, the Bott index $\mathcal{B}_{\mathcal{N}}$ is given by~\cite{PhysRevLett.125.217202,huang2018theory}
\begin{equation}
	\begin{split}
		\mathcal{B}_{\mathcal{N}}=-\frac{1}{2\pi}\mathrm{Im}\left\{\mathrm{Tr}\left[\log(VUV^\dagger U^\dagger)\right]\right\}.
	\end{split}
	\label{eq:P4}
\end{equation}
Using Jacobi's formula, $\log(\det A)=\mathrm{Tr}(\log A)$, we can relate the Bott index to the determinant of the matrix $VUV^{\dagger}U^{\dagger}$. Since det $VUV^{\dagger}U^{\dagger}$ must be real and of the form $e^{2\pi mi+r}$, where $m$ is an integer and $r$ is a real number, the Bott index ${\mathcal{B}}_{\mathcal{N}}$ is quantized to integer values provided that $U$ and $V$ are full-rank matrices. The Bott index of the many-body ground state is given by $\mathcal{B}=\sum_{\mathcal{N}\in\mathrm{occ.}} {\mathcal{B}}_{\mathcal{N}}$, where ${\rm occ.}$ denotes the occupied bands of the interacting fermions in the Lieb lattice.

\subsection{Spectral function}\label{section:Spectral
	function}
To further characterize the topological nature of the many-body ground state, we employ spectral function analysis through retarted Green's function. This approach provides direct access to edge states, which serve as smoking-gun evidence of nontrivial topology. The spectral function is given by~\cite{PhysRevLett.84.522,PhysRevB.94.214510,vasic2015chiral}  
\begin{equation}  
	A(\omega,\boldsymbol{k}) = -\frac{1}{\pi}\mathrm{Tr}\left[\mathrm{Im}\,G^{\mathrm{R}}(\omega+i\delta,\boldsymbol{k})\right],
	\label{eq:spectrum}  
\end{equation}  
where $\delta$ is the inverse lifetime of the quasiparticle. Our DMFT implementation in cylindrical boundary conditions unambiguously demonstrates bulk-boundary correspondence. As shown in Figs. 2(e) and (f) in the main text, the edge spectral weight appears across the bulk gap only for open boundary conditions, signaling topologically protected edge states.
Through a combination of diverse methods, we have achieved a comprehensive understanding of topological properties of the interacting fermionic system, and reinforced the validity of our results.

\section{Magnetic order parameter}\label{chapter:mag invariant}

In this part, we provide more information about magnetic order parameters and phase transitions. To characterize the rich magnetic phases, local order parameters, including spin order parameter $\langle \mathbf{S}\rangle$, spin structure factor $\mathbf{S}_q$, and charge modulation amplitudes $\delta_n$ are introduced, where 
\begin{equation}
	\begin{split}
		&\mathbf{S}^x =\frac{1}{2}(c_{\uparrow}^{\dagger}c_{\downarrow}+c_{\downarrow}^{\dagger}c_{\uparrow})\\
		&\mathbf{S}^y =\frac{1}{2}(-i c_{\uparrow}^{\dagger}c_{\downarrow}+i c_{\downarrow}^{\dagger}c_{\uparrow})\\
		&\mathbf{S}^z = \frac{1}{2}(n_{\uparrow}-n_{\downarrow})\\
		&\mathbf{S}_q = \frac{1}{N}\left|\sum_{\boldsymbol{r}} \langle\mathbf{S}_{\boldsymbol{r}}\rangle e^{i\mathbf{q} \cdot \boldsymbol{r}}\right|\\
		&\delta_n = n_{\boldsymbol{r},A} - n_{\boldsymbol{r}',A},
	\end{split}
	\label{eq:mag order}
\end{equation}
with $N$ denoting the number of unit cell.

\begin{figure*}[htbp]
	\centering
	\includegraphics[ width=0.98\textwidth]{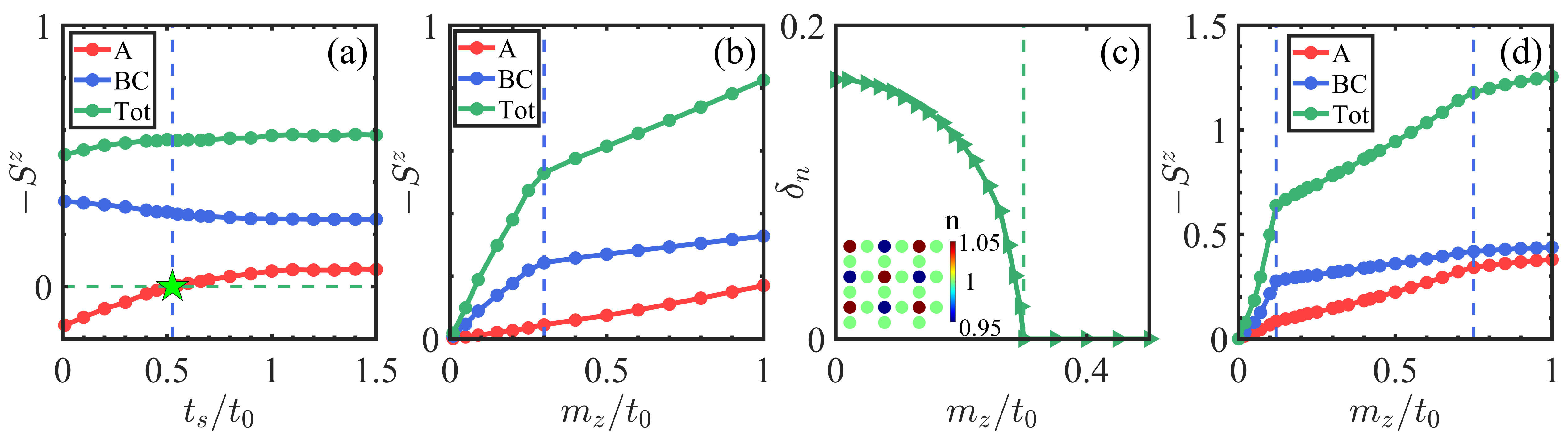} 
	\caption{Magnetic order parameters calculated using DMFT. (a)(b) Spin order parameter $\mathbf{\langle S}^z\rangle$ as a function of (a) $t_s$ for $U=3t_{0}$ and $m_z=0.4t_{0}$, and (b) $m_z$ for  $U=3t_{0}$ and $t_s=1.1t_{0}$. (c) Charge modulation amplitudes $\delta_n$ versus $m_z$ for $U=3t_{0}$ and $t_s=1.1t_{0}$, with the inset illustrating the real-space density distributions. (d) Spin order parameters $\mathbf{\langle S}^z\rangle$ for $U=8t_{0}$ and $t_s=1.5t_{0}$.}
	\label{fig:SM_order}
\end{figure*}

Based on these order parameters, the magnetic phase diagrams can be obtained, as shown in Fig. 3(a)(b) in the main text. For $U=3t_{0}$ [Fig. 3(a)], the system stabilizes a ferrimagnetic ($z$-FIM) phase when $m_z < t_{0}$ and $t_s < t_{0}$. The ferrimagnetic nature is confirmed by staggered sublattice spin polarization $|\langle \mathbf{S}^z\rangle|$, with $\langle \mathbf{S}^z_{\text{BC}}\rangle \neq \langle \mathbf{S}^z_{\text{A}}\rangle$ in both magnitude and sign, as illustrated in Fig.~\ref{fig:SM_order}(a), where $\langle \mathbf{S}^z_{\rm Tot}\rangle\equiv\langle{\bf S}^z_{\bf A}\rangle + \langle{\bf S}^z_{\bf B}\rangle + \langle{\bf S}^z_{\bf C}\rangle$ for each unit cell. Increasing $t_s$ drives $\langle {\bf S}^z\rangle$ of the A-sublattice to undergo a sign reversal [green star in Fig.~\ref{fig:SM_order}(a)], signaling a crossover from the $z$-FIM to the partially magnetic (PM) phase. In the SOC dominant regime ($t_s \approx t_{0}$), a vortex (Vx) phase emerges [Fig. 3(a)]. Interestingly, we observe charge modulations for the $A$-site in the Vx phase [inset of Fig.~\ref{fig:SM_order}(c)]. In addition, the Vx-to-PM transition is identified by abrupt changes in $BC$-site $\langle \mathbf{S}^z \rangle$ [Fig.~\ref{fig:SM_order}(b)] and $A$-site charge modulations [Fig.~\ref{fig:SM_order}(c)].

In the strongly interacting regime, the order parameter $\langle {\bf S}^z\rangle$ remains effective in distinguishing magnetic phases. For $U=8t_0$ [Fig. 3(b) in the main text], four quantum phases are predicted, including the ferrimagnetic ($z$-FIM), spin-spiral (SSM), partially (PM), and saturated (STM) phases. An example of the phase boundaries is given in Fig.~\ref{fig:SM_order}(d) as function of $m_z$. For fixed $t_s=1.5t_0$,  $\langle {\bf S}^z\rangle$ demonstrates two sharp changes at the critical points: $m_z\approx0.12t_0$ for the SSM-to-PM transition, and $m_z\approx0.75t_0$ for the PM-to-STM transition.

\section{Self energy analysis for the topology of interacting systems}\label{chapter:self energy}

\begin{figure}[htbp]
	\centering
\includegraphics[width=0.645\textwidth]{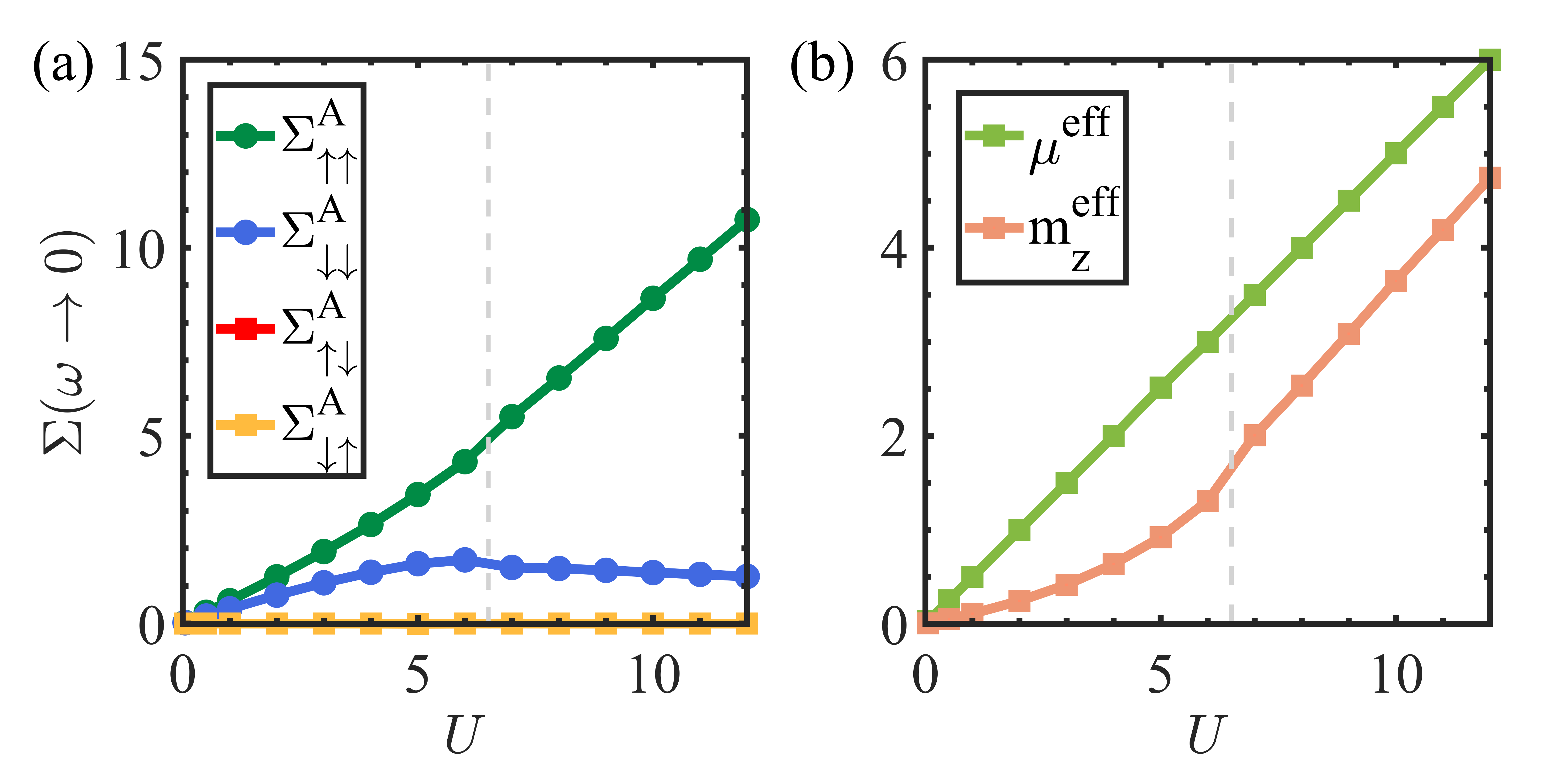} 
	\caption{Self-energy of the interacting fermions in the Lieb lattice. (a) Self-energies as a function of interactions $U$ for the $A$-site, obtained by DMFT. Diagonal terms of the local self-energy dominate, while the off-diagonal ones are suppressed. (b) Self-energies in (a) are decomposed as the effective magnetic field $m_z^{\rm eff}\equiv \frac{\Sigma_{\uparrow\uparrow}-\Sigma_{\downarrow\downarrow}}{2}$, and the effective chemical potential $\mu^{\rm eff} \equiv \frac{\Sigma_{\uparrow\uparrow}+\Sigma_{\downarrow\downarrow}}{2}$. The parameters are set to be $m_z=1.0t_{0}$ and $t_s=1.5t_{0}$.
	}	\label{fig:self}
\end{figure}
In this part, we provide the underlying mechanics of topology of the interacting fermionic systems. Within DMFT, topological properties of the interacting fermionic system are obationed via the Bott index $\mathcal B$. As shown in Eq.~(\ref{Bott_selfenergy}), the crucial point for calculating topological invariant $\mathcal B$ is to obtain the self energy of the interacting system. 

Using DMFT, we obtain the self-energy $\Sigma_{\nu\nu^\prime}$ of the interacting system in a two-dimensional spin-orbit-coupled Lieb lattice. Analysis reveals that the self-energy is dominated by the diagonal terms $\Sigma_{\nu\nu}$, where the off-diagonal terms $\Sigma_{\uparrow\downarrow}$ are approximately zero. One typical example is provided in Fig.~\ref{fig:self}(a)(b) to illustrate the dependence of the self-energy on interactions $U$. Considering the self-energy components are positive for all lattice sites, the diagonal terms can be described by an effective structure of the form $\Sigma_{\mathrm{eff}}=\frac{\Sigma_{\uparrow\uparrow}-\Sigma_{\downarrow\downarrow}}{2} \sigma_z+\frac{\Sigma_{\uparrow\uparrow}+\Sigma_{\downarrow\downarrow}}{2} \sigma_0 $. The first term acts as an effective magnetic field $m_z^{\rm eff}$, nonmonotonically manipulating topological properties of the system.
The second term can be neglected in constructing the Bott index, since it only provides an energy shift in the eigenvalues, unrelated to topological properties of the interacting system. Thus, the influence of the interaction on topology is akin to the effective magnetic field.  %

\section{Experimental realization}\label{chapter:Experimental realization}
In this section, we detail the experimental realization of a spin-orbit-coupled Lieb lattice in ultracold atoms. The experimental setup is shown in Fig.~5(a) in the main text. The optical configuration employs retro-reflected laser beams to generate standing-wave fields:
\begin{align}
	\begin{split}
		&{\bf E}_{1x}=\hat{e}_yE_{1x}e^{i (\alpha+\alpha_L/2)}\cos{(k_0x-\alpha_L/2)},\quad {\bf E}_{1y}=\hat{e}_xE_{1y}e^{i (\beta+\beta_L/2)}\cos{(k_0y-\beta_L/2)},\\
		&{\bf E}_{2x}=\hat{e}_yE_{2x}e^{i (\alpha^{'}+\alpha_L^{'}/2)}\cos{(k_0x/2-\alpha_L^{'}/2)},\quad {\bf E}_{2y}=\hat{e}_xE_{2y}e^{i (\beta^{'}+\beta_L^{'}/2)}\cos{(k_0y/2-\beta_L^{'}/2)},\\
		&{\bf E}_{3}={\bf E}_{3x}+{\bf E}_{3y}=\hat{e}_ze^{i (\alpha^{''}+\alpha_L^{''}/2)}\left[E_{3x}\cos{(k_0x-\alpha_L^{''}/2)}+E_{3y}\cos{(k_0y-\alpha_L^{''}/2)}\right],
	\end{split}
\end{align}
where $E_{n\mu}$ denote the amplitude of the $n$-th beam propagating along the $\mu$-direction, $\alpha,\beta,\alpha^{'},\beta^{'},\alpha^{''}$ are the initial phases, and $\alpha_L,\beta_L,\alpha_L^{'},\beta_L^{'},\alpha_L^{''}$ are the phases acquired by the reflective optical path. Here, in order to realize the Lieb lattice, we set the phase as follows:
\begin{align}
	\begin{split}
		&\alpha=\alpha^{'}=\alpha^{''}=\beta=\beta^{'},\quad \alpha_L=\beta_L,\\
		&\alpha_L^{'}=\alpha_L+\pi/2,\quad\beta_L^{'}=\beta_L-\pi/2,\\
		&\alpha_L^{''}=\alpha_L-\pi,
	\end{split}
\end{align}
which can be precisely controlled via electro-optic modulators or using $\lambda/2$ or $\lambda/4$ waveplates. With these phase configurations and the condition $E_{3x}=E_{3y}=E_{3}$, we rewrite the beam fields as
\begin{align}
	\begin{split}
		&{\bf E}_{1x}=\hat{e}_yE_{1x}e^{i (\alpha+\alpha_L/2)}\cos{(k_0x-\alpha_L/2)},\quad {\bf E}_{1y}=\hat{e}_xE_{1y}e^{i (\alpha+\alpha_L/2)}\cos{(k_0y-\alpha_L/2)},\\
		&{\bf E}_{2x}=\hat{e}_yE_{2x}e^{i (\alpha+\alpha_L/2+\pi/4)}\cos{(k_0x/2-\alpha_L/2-\pi/4)},\,\, {\bf E}_{2y}=\hat{e}_xE_{2y}e^{i (\alpha+\alpha_L/2-\pi/4)}\cos{(k_0y/2-\alpha_L/2+\pi/4)},\\
		&{\bf E}_{3}=\hat{e}_zE_3e^{i (\alpha+\alpha_L/2-\pi/2)}\left[\cos{(k_0x-\alpha_L/2+\pi/2)}+\cos{(k_0y-\alpha_L/2+\pi/2)}\right].
	\end{split}
\end{align}
In the following, we take $\prescript{40}{}{\rm K}$ atoms as an example to illustrate our scheme, while our results are generally applicable to other atomic species. 

\subsection{Lattice and Raman potentials}\label{section:Lattice and Raman potentials}

\begin{figure}
	\includegraphics[width=0.98\textwidth]{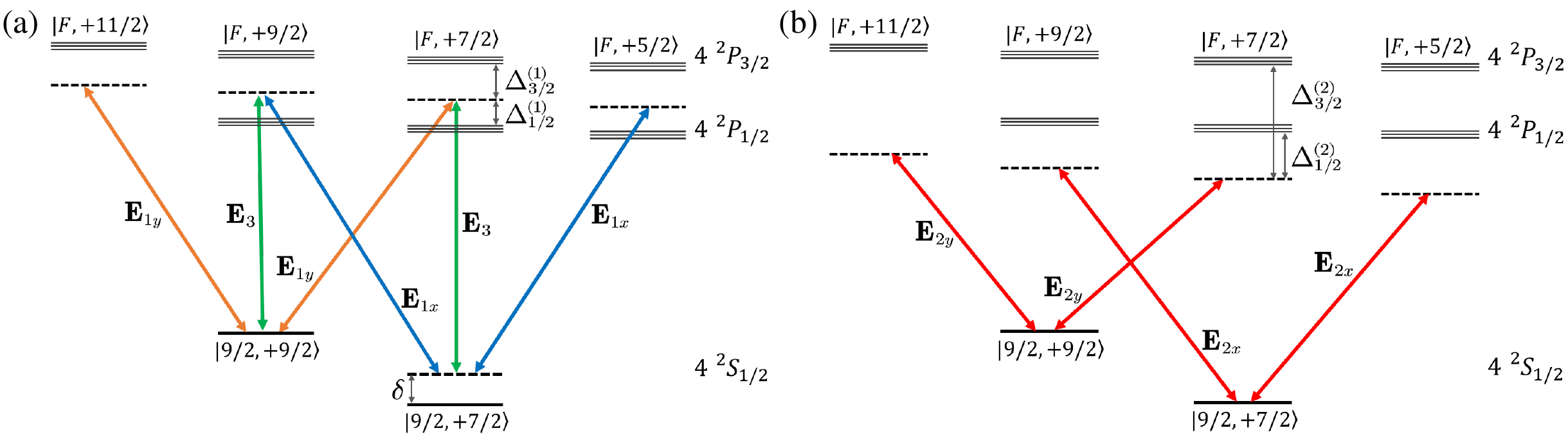}
	\caption{Laser couplings for $\prescript{40}{}{\rm K}$ atoms. (a) Optical transitions induced by beams ${\bf E}_{1x,1y}$ and ${\bf E}_{3}$. (b) Optical transitions induced by beams ${\bf E}_{2x,2y}$.
	}\label{fig:SM_optical}
\end{figure}

As shown in Fig.~\ref{fig:SM_optical}, the spin-$1/2$ system is constructed by selecting $\ket{\uparrow}=\ket{F=9/2, m_F=+9/2}$ and $\ket{\downarrow}=\ket{9/2,+7/2}$. When incorporating contributions from both $D_1$ $(4\prescript{2}{}{S}_{1/2}\rightarrow4\prescript{2}{}{P}_{1/2})$ and $D_2$ $(4\prescript{2}{}{S}_{1/2}\rightarrow4\prescript{2}{}{P}_{3/2})$ transitions, the optical lattice potentials are given by ($\sigma=\uparrow,\downarrow$)
\begin{align}
	V_{\sigma}=\sum_{F,\nu}\frac{1}{\Delta_{3/2}}\left(\left|\Omega_{\sigma F,\nu z}^{(3/2)}\right|^2+\left|\Omega_{\sigma F,\nu +}^{(3/2)}\right|^2+\left|\Omega_{\sigma F,\nu-}^{(3/2)}\right|^2\right)+\sum_{F,\nu}\frac{1}{\Delta_{1/2}}\left(\left|\Omega_{\sigma F,\nu z}^{(1/2)}\right|^2+\left|\Omega_{\sigma F,\nu+}^{(1/2)}\right|^2+\left|\Omega_{\sigma F,\nu -}^{(1/2)}\right|^2\right),
\end{align}
where $\Delta_{J}=\Delta^{(1)}_{J}$ for couplings of ${\bf E}_{1x,1y}$ and ${\bf E}_{3}$, $\Delta_{J}=\Delta^{(2)}_{J}$ for couplings of ${\bf E}_{2x,2y}$, $\Omega_{\sigma F,\nu z}^{(J)}=\matrixel{\sigma}{er}{F,m_{F\sigma},J}\hat{e}_{z}\cdot {\bf E}_{\nu}$ for $\pi$-transitions, and $\Omega_{\sigma F,\nu\pm}^{(J)}=\matrixel{\sigma}{er}{F,m_{F\sigma}\pm1,J}\hat{e}_{\pm}\cdot {\bf E}_{\nu}$ for $\sigma^{\pm}$-transitions, with $J=1/2,3/2$ and $\nu=1x,1y,2x,2y,3$. 
Here, $\hat{e}_{+}=(\hat{x}-i \hat{y})/\sqrt{2}$ and $\hat{e}_{-}=(\hat{x}+i \hat{y})/\sqrt{2}$. From the data of dipole matrix elements of $\prescript{40}{}{\rm K}$, 
we can obtain the spin-independent optical potentials
\begin{align}
	\begin{split}
		V_{\uparrow}=V_{\downarrow}=&\left[ V_{1x} \cos^2 \left( k_0 x - \frac{\alpha_L}{2} \right) + V_{1y} \cos^2 \left( k_0 y - \frac{\alpha_L}{2} \right) \right] \\
		&+ \left[ V_{2x} \cos^2 \left( \frac{k_0 x}{2}  - \frac{\pi}{4} - \frac{\alpha_L}{2} \right) + V_{2y} \cos^2 \left( \frac{k_0 y}{2} + \frac{\pi}{4} - \frac{\alpha_L}{2} \right) \right] \\
		&+ V_{3} \left[ \sin \left( k_0 x - \frac{\alpha_L}{2} \right) + \sin \left( k_0 y - \frac{\alpha_L}{2} \right) \right]^2,
	\end{split}
\end{align}
where $V_{\nu}=\frac{\alpha_{D_1}^2}{3}\left(\frac{1}{\Delta_{1/2}}+\frac{2}{\Delta_{3/2}}\right)E_{\nu}^2$ for $\nu=1x,1y,2x,2y,3$.
Here, the transition matrix elements $\alpha_{D_1}\equiv\langle J=1/2||e{\bf r}||J'=1/2\rangle$, $\alpha_{D_2}\equiv\langle J=1/2||e{\bf r}||J'=3/2\rangle$ and
$\alpha_{D_2}\approx\sqrt{2}\alpha_{D_1}$.
Meanwhile, two Raman couplings are generated respectively by the pairs $({\bf E}_{3},{\bf E}_{1x})$ and $({\bf E}_{1y},{\bf E}_{3})$ [see Fig.~\ref{fig:SM_optical}(a)], which requires the frequencies to satisfy the condition $\omega_{1x}-\omega_{3}=\omega_{3}-\omega_{1y}$ and leads to the potentials
\begin{align}
	\begin{split}
		M_1 &= \sum_{F} \frac{\Omega_{\uparrow F,3z}^{(3/2)*} \cdot \Omega_{\downarrow F,1x+}^{(3/2)}}{\Delta^{(1)}_{3/2}} + \sum_{F} \frac{\Omega_{\uparrow F,3z}^{(1/2)*} \cdot \Omega_{\downarrow F,1x+}^{(1/2)}}{\Delta^{(1)}_{1/2}}, \\
		M_2 &= \sum_{F} \frac{\Omega_{\uparrow F,1y-}^{(3/2)*} \cdot \Omega_{\downarrow F,3z}^{(3/2)}}{\Delta^{(1)}_{3/2}} + \sum_{F} \frac{\Omega_{\uparrow F,1y-}^{(1/2)*} \cdot \Omega_{\downarrow F,3z}^{(1/2)}}{\Delta^{(1)}_{1/2}}.
	\end{split}
\end{align}
We thus have
\begin{align}
	\begin{split}
		M_1 &= M_{01}\left[\cos(k_0 x - \frac{\alpha_L}{2}) \sin(k_0 x - \frac{\alpha_L}{2})+ \cos(k_0 x-\frac{\alpha_L}{2}) \sin(k_0 y-\frac{\alpha_L}{2})\right], \\
		M_2 &= i M_{02}\left[\sin(k_0 x - \frac{\alpha_L}{2})\cos(k_0 y -\frac{\alpha_L}{2})+\sin(k_0 y - \frac{\alpha_L}{2})\cos(k_0 y - \frac{\alpha_L}{2})\right],
	\end{split}
\end{align}
where 
\begin{align}
	M_{01/02} =\frac{\alpha_{D_1}^2}{9}\left( \frac{1}{\Delta^{(1)}_{1/2}}-\frac{1}{\Delta^{(1)}_{3/2}}\right)E_{3}E_{1x/1y}.
\end{align}
Note that the term $\cos(k_0 x-\alpha_L/2) \sin(k_0 x-\alpha_L/2)$ [analogously for $\sin(k_0 y-\alpha_L/2) \cos(k_0 y-\alpha_L/2)$] is antisymmetric about both lattice sites and nearest-neighbor bond centers in the $x$ direction, thus contributing negligibly to lowest $s$-band physics in the tight-binding approximation due to symmetry constraints on on-site and nearest-neighbor integrals. Neglecting such terms finally yields
\begin{align}
	\begin{split}
		M_1 &= M_{01} \cos(k_0 x-\frac{\alpha_L}{2}) \sin(k_0 y-\frac{\alpha_L}{2}), \\
		M_2 &= i M_{02}\sin(k_0 x - \frac{\alpha_L}{2})\cos(k_0 y -\frac{\alpha_L}{2}).
	\end{split}
\end{align}
This result indicates that the effective Raman couplings emerge only from orthogonal polarization pairs $({\bf E}_{3y}, {\bf E}_{1x})$ and $({\bf E}_{1y}, {\bf E}_{3x})$.

The total Hamiltonian for the two-dimentional spin-orbit-coupled Lieb lattice then reads
\begin{align}~\label{H_S}
	H=\left[\frac{{\bf k}^2}{2m}+V_{\rm lat}({\bf r})\right]\otimes {\bf 1}+\mathcal{M}_1({\bf r})\sigma_x+\mathcal{M}_2({\bf r})\sigma_y+m_z\sigma_z, 
\end{align}
where the Lieb lattice potential
\begin{align}
	V_{\rm lat}({\bf r})=V_1\left[\cos^2(k_0 x)+\cos^2(k_0 y)\right]+V_2\left[\cos^2\left(\frac{k_0 x}{2}-\frac{\pi}{4}\right)+\cos^2\left(\frac{k_0 y}{2}+\frac{\pi}{4}\right)\right]+V_3\left[\sin\left(k_0 x\right)+\sin\left(k_0 y\right)\right]^2,
\end{align}
Raman coupling potentials
\begin{align}
	\mathcal{M}_1({\bf r})=M_0\cos(k_0 x)\sin(k_0 y), \quad \mathcal{M}_2({\bf r})=M_0\sin(k_0 x)\cos(k_0 y),
\end{align}
and the Zeeman constant $m_z=\delta/2$.
Here, we have assumed $\alpha_L =0$, $V_{1x}=V_{1y}=V_{1}$, $V_{2x}=V_{2y}=V_{2}$ ,and $M_{01}=M_{02}=M_{0}$ for simplicity. 

The diagonalization of the Hamiltonian~\eqref{H_S} can be achieved by constructing a complete set (basis) of plane waves
$\{\psi_{m,n}^{\uparrow}(\mathbf{k}), \psi_{p,q}^{\downarrow}(\mathbf{k})\}$, where
\begin{align}
	&\psi_{m,n}^{\uparrow}(\mathbf{k})=\frac{1}{\sqrt{S}} e^{i(k_x+mk_0)x} e^{i(k_y+nk_0)y}, \\
	&\psi_{p,q}^{\downarrow}(\mathbf{k})=\frac{1}{\sqrt{S}} e^{i(k_x+pk_0+k_0)x} e^{i(k_y+qk_0+k_0)y}.
\end{align}
Here, $m$, $n$, $p$ and $q$ are integers, and $S$ denotes the system area. Using the relation $\langle\psi_{m',n'}^{\sigma}|\psi_{m,n}^{\sigma'}\rangle=\delta_{m',m}\delta_{n',n}\delta_{\sigma\sigma'}$, one can write the Hamiltonian in the matrix form.

\subsection{Tight-binding model}\label{section:Tight-binding model}

We consider the lowest $s$-orbitals at each lattice site and include only nearest-neighbor hopping terms. The tight-binding Hamiltonian generically takes the form 
\begin{align}
	H_{\rm TB}=\sum_{<\vec{i},\vec{j}>,\sigma}t_0^{\vec{i} \vec{j}}c_{\vec{i}\sigma}^{\dagger}c_{\vec{j}\sigma}+\sum_{<\vec{i},\vec{j}>}t_{\rm s}^{\vec{i} \vec{j}}(c_{\vec{i}\uparrow}^{\dagger}c_{\vec{j}\downarrow}+c_{\vec{i}\downarrow}^{\dagger}c_{\vec{j}\uparrow}+{\rm h.c.})+\sum_{\vec{i}}m_z(n_{\vec{i}\uparrow}-n_{\vec{i}\downarrow}),
\end{align}
where $\vec{i}\equiv(i_x,i_y)$ denotes the lattice site indices and the particle number operators $n_{\vec{i}\sigma}=c_{\vec{i}\sigma}^\dag c_{\vec{i}\sigma}$.

The spin-conserved hopping amplitude $t_0^{\vec{i}\vec{j}}$, induced by the lattice potential $V_{\rm lat}$, exhibits identical magnitude along both $x$- and $y$-directions due to lattice symmetry, yielding $t_0^{\vec{i}\vec{i}\pm\vec{1}}=-t_0$, with
\begin{align}
	t_0=-\int d^2{\bf r}\phi_{s}^{A\ast}(x,y)\left[\frac{{\bf k}^2}{2m}+V_{\rm lat}({\bf r})\right]\phi_{s}^B(x-a,y),
\end{align}
where $a=\pi/k_0$, and $\phi_{s}^A$ ($\phi_{s}^B$) denotes the spin-independent $s$-orbital Wannier function at a sublattice site $A$ ($B$).
The spin-flip hopping amplitude is given by
\begin{align}~\label{tsij}
	t_{\rm s}^{\vec{i} \vec{j}}=\int d^2{\bf r}\phi_{s}^{\vec{i}\ast}({\bf r})\left[\mathcal{M}_1({\bf r})-i\mathcal{M}_2({\bf r})\right]\phi_{s}^{\vec{j}}({\bf r}).
\end{align}
Owing to the bipartite nature of the Lieb lattice, nearest-neighbor hoppings between sublattices $A$ and $B$ (or $C$) require separate treatment. Without loss of generality, we fix the coordinate origin at a sublattice $A$ site and set $V_1>0$. When indices $i_x$ and $i_y$ are both even, Eq.~\eqref{tsij} then explicitly describes a spin-flip hopping process $B/C\to A$ between nearest-neighbor sites.
The corresponding hopping amplitude induced by $\mathcal{M}_1({\bf r})$ is
\begin{align}~\label{tsAB}
	\begin{split}
		\int d^2{\bf r}\phi_{s}^{\vec{i}\ast}({\bf r})\mathcal{M}_1({\bf r})\phi_{s}^{\vec{j}}({\bf r})&=M_0\int d^2{\bf r}\phi_{s}^{A\ast}({\bf r}-{\bf r}_{\vec{i}})\cos(k_0 x)\sin(k_0 y)\phi_{s}^B({\bf r}-{\bf r}_{\vec{j}})\\
		&=M_0\int d^2{\bf r}\phi_{s}^{A\ast}({\bf r})\cos(k_0 x+i_x\pi)\sin(k_0 y+i_y\pi)\phi_{s}^B({\bf r}-{\bf r}_{\vec{j}}+{\bf r}_{\vec{i}})\\
		&=(-1)^{i_x+i_y}M_0\int d^2{\bf r}\phi_{s}^{A\ast}({\bf r})\cos(k_0 x)\sin(k_0 y)\phi_{s}^B({\bf r}-{\bf r}_{\vec{j}}+{\bf r}_{\vec{i}}),
	\end{split}
\end{align}
where the coefficient $(-1)^{i_x+i_y}=1$ is determined by the {\it terminal} sublattice $A$ site’s coordinates $(i_x, i_y)$ in the spin-flip hopping process. 
When either $i_x$ or $i_y$ (but not both) is odd, Eq.~\eqref{tsij} governs a spin-flip hopping process $A\to B/C$, with the $\mathcal{M}_1$-induced hopping amplitude given by
\begin{align}~\label{tsBA}
	\begin{split}
		\int d^2{\bf r}\phi_{s}^{\vec{i}\ast}({\bf r})\mathcal{M}_1({\bf r})\phi_{s}^{\vec{j}}({\bf r})
		&=M_0\int d^2{\bf r}\phi_{s}^{B\ast}({\bf r}-{\bf r}_{\vec{i}})\cos(k_0 x)\sin(k_0 y)\phi_{s}^A({\bf r}-{\bf r}_{\vec{j}})\\
		&=(-1)^{i_x+i_y}M_0\int d^2{\bf r}\phi_{s}^{B\ast}({\bf r})\cos(k_0 x)\sin(k_0 y)\phi_{s}^A({\bf r}-{\bf r}_{\vec{j}}+{\bf r}_{\vec{i}}),
	\end{split}
\end{align}
where $(-1)^{i_x+i_y}=-1$ also arises from the odd parity of the {\it terminal} sublattice $B$ site’s coordinates $(i_x, i_y)$.
Similarly, for the Raman potential $\mathcal{M}_2$, we have
\begin{align}
	\int d^2{\bf r}\phi_{s}^{\vec{i}\ast}({\bf r})\mathcal{M}_2({\bf r})\phi_{s}^{\vec{j}}({\bf r})
	=(-1)^{i_x+i_y}M_0\int d^2{\bf r}\phi_{s}^{A\ast}({\bf r})\sin(k_0 x)\cos(k_0 y)\phi_{s}^C({\bf r}-{\bf r}_{\vec{j}}+{\bf r}_{\vec{i}})
\end{align}
for the case that both $i_x$ and $i_y$ are even, and
\begin{align}
	\int d^2{\bf r}\phi_{s}^{\vec{i}\ast}({\bf r})\mathcal{M}_2({\bf r})\phi_{s}^{\vec{j}}({\bf r})
	&=(-1)^{i_x+i_y}M_0\int d^2{\bf r}\phi_{s}^{C\ast}({\bf r})\sin(k_0 x)\cos(k_0 y)\phi_{s}^A({\bf r}-{\bf r}_{\vec{j}}+{\bf r}_{\vec{i}}),
\end{align}
corresponding to the case that either $i_x$ or $i_y$ is odd.
Note that the maximally localized Wannier functions $\phi_{s}^{\vec{i}}({\bf r})$
are real-valued and exhibit site-centered symmetry: $\phi_{s}^{\vec{i}}({\bf r})=\phi_{s}^{\vec{i}}(-{\bf r})$. Furthermore, when $V_1>0$, 
the Raman potential $\mathcal{M}_1$ ($\mathcal{M}_2$) preserves bond-centered symmetry between coupled adjacent sites in the $x$ ($y$)-direction. 
These symmetries lead to
\begin{align}
	\begin{split}
		&\int d^2{\bf r}\phi_{s}^{A\ast}(x,y)\cos(k_0 x)\sin(k_0 y)\phi_{s}^B(x-a,y)=\int d^2{\bf r}\phi_{s}^{B\ast}(x,y)\cos(k_0 x)\sin(k_0 y)\phi_{s}^A(x-a,y)\\
		&=\int d^2{\bf r}\phi_{s}^{A\ast}(x,y)\sin(k_0 x)\cos(k_0 y)\phi_{s}^C(x,y-a)=\int d^2{\bf r}\phi_{s}^{C\ast}(x,y)\sin(k_0 x)\cos(k_0 y)\phi_{s}^A(x,y-a).
	\end{split}
\end{align}
and 
\begin{align}
	\int d^2{\bf r}\phi_{s}^{A\ast}(x,y)\cos(k_0 x)\sin(k_0 y)\phi_{s}^B(x+a,y)=-\int d^2{\bf r}\phi_{s}^{A\ast}(x,y)\cos(k_0 x)\sin(k_0 y)\phi_{s}^B(x-a,y).
\end{align}
We then define
\begin{align}
	t_{\rm s}=M_0\int d^2{\bf r}\phi_{s}^{A\ast}(x,y)\cos(k_0 x)\sin(k_0 y)\phi_{s}^B(x-a,y),
\end{align}
and have the following results:
\begin{align}
	t_{\rm s}^{i_x,i_x\pm1}=\pm(-1)^{i_x+i_y}t_{\rm s}, \quad  t_{\rm s}^{i_y,i_y\pm1}=\pm i(-1)^{i_x+i_y}t_{\rm s}.
\end{align}

Finally, the total Hamiltonian can be rewritten as
\begin{align}
	\begin{split}
		H_{\rm TB}=&-t_0\sum_{<\vec{i},\vec{j}>,\sigma}c_{\vec{i}\sigma}^{\dagger}c_{\vec{j}\sigma}+\left[\sum_{\vec{i}\in A,B}(-1)^{i_x+i_y}t_{\rm s}
		\left(c_{i_x\uparrow}^{\dagger}c_{i_x+1\downarrow}-c_{i_x\uparrow}^{\dagger}c_{i_x-1\downarrow}\right)+{\rm h.c.}\right]\\
		&+\left[\sum_{\vec{i}\in A,C}(-1)^{i_x+i_y}i t_{\rm s}\left(c_{i_y\uparrow}^{\dagger}c_{i_y+1\downarrow}-c_{i_y\uparrow}^{\dagger}c_{i_y-1\downarrow}\right)+{\rm h.c.}\right]
		+\sum_{\vec{i}\in A,B,C}m_z\left(n_{\vec{i}\uparrow}-n_{\vec{i}\downarrow}\right).\\
	\end{split}
\end{align}
We further do the transformation for spin-down operators as $c_{\vec{i}\downarrow}\rightarrow (-1)^{i_x+i_y-1}c_{\vec{i}\downarrow}$, which recasts the Hamiltonian into
\begin{align}
	\begin{split}
		H_{\rm TB}=&-t_0\sum_{<\vec{i},\vec{j}>}\left(c_{\vec{i}\uparrow}^{\dagger}c_{\vec{j}\uparrow}-c_{\vec{i}\downarrow}^{\dagger}c_{\vec{j}\downarrow}\right)+\left[\sum_{\vec{i}\in A,B}t_{\rm s}
		\left(c_{i_x\uparrow}^{\dagger}c_{i_x+1\downarrow}-c_{i_x\uparrow}^{\dagger}c_{i_x-1\downarrow}\right)+{\rm h.c.}\right]\\
		&+\left[\sum_{\vec{i}\in A,C}i t_{\rm s}\left(c_{i_y\uparrow}^{\dagger}c_{i_y+1\downarrow}-c_{i_y\uparrow}^{\dagger}c_{i_y-1\downarrow}\right)+{\rm h.c.}\right]
		+\sum_{\vec{i}\in A,B,C}m_z\left(n_{\vec{i}\uparrow}-n_{\vec{i}\downarrow}\right).\\
	\end{split}
\end{align}

\end{document}